\documentclass{article}

\usepackage{arxiv}

\usepackage[utf8]{inputenc} 
\usepackage[T1]{fontenc}    
\usepackage{hyperref}       
\usepackage{url}            
\usepackage{booktabs}       
\usepackage{amsfonts}       
\usepackage{nicefrac}       
\usepackage{microtype}      
\usepackage{lipsum}         
\usepackage{graphicx}
\usepackage{natbib}
\usepackage{doi}

\usepackage{amsmath}
\usepackage{subcaption}
\usepackage{multirow}
\usepackage{ltablex}
\usepackage{longtable}
\usepackage{enumitem}
\usepackage{xspace}
\usepackage{array}
\usepackage{verbatim}
\usepackage{color, colortbl}
\setenumerate{partopsep=0cm, topsep=0cm}
\newcommand{\matr}[1]{\mathbf{#1}}

\newcommand{\jitbl}{JINGO\xspace}
\newcommand{\jitblshort}{JINGO\xspace}
\newcommand{\corleyshort}{Corley\xspace}
\newcommand{\argmin}{\mathop{\mathrm{arg\,min}}}

\definecolor{MRR}{rgb}{0.85,0.95,1}
\definecolor{MAP}{rgb}{0.7,0.9,0.7}
\definecolor{gray}{rgb}{0.9, 0.9, 0.9}
\definecolor{ggray}{rgb}{0.77, 0.77, 0.77}
\newcolumntype{L}{>{\raggedleft\arraybackslash}m{0.7cm}}
\newcolumntype{W}{>{\raggedleft\arraybackslash}m{1.1cm}}

\title{Online Adaptable Bug Localization for Rapidly Evolving Software}

\date{}

\author{Agnieszka Ciborowska \\
	Department of Computer Science\\
	Virginia Commonwealth University\\
	Richmond, VA, USA \\
	\texttt{ciborowskaa@vcu.edu} \\
	\And
	Michael J. Decker \\
	Department of Computer Science\\
	Bowling Green State University\\
	Bowling Green, OH, USA \\
	\texttt{mdecke@bgsu.edu} \\
	\And
	Kostadin Damevski \\
	Department of Computer Science\\
	Virginia Commonwealth University\\
	Richmond, VA, USA \\
	\texttt{kdamevski@vcu.edu} \\
}

\begin{document}
\maketitle

\begin{abstract}
Bug localization aims to reduce debugging time by recommending program elements that are relevant for a specific bug report. To date, researchers have primarily addressed this problem by applying different information retrieval techniques that leverage similarities between a given bug report and source code. However, with modern software development trending towards increased speed of software change and continuous delivery to the user, the current generation of bug localization techniques, which cannot quickly adapt to the latest version of the software, is becoming inadequate. In this paper, we propose a technique for  online bug localization, which enables rapidly updatable bug localization models.
More specifically, we propose a streaming bug localization technique, based on an ensemble of online topic models, that is able to adapt to both specific (with explicit code mentions) and more abstract bug reports. By using changesets (diffs) as the input instead of a snapshot of the source code, the model naturally integrates defect prediction and co-change information into its prediction.
Initial results indicate that the proposed approach improves bug localization performance for 42 out of 56 evaluation projects, with an average MAP improvement of 5.9\%.
\end{abstract}


\section{Introduction}

Locating and fixing software bugs has persisted as one of the most common and important tasks software developers face on a daily basis. In well organized software projects, all bug occurrences are reported and preserved in an issue tracker, while the resulting software modifications are stored in a version control system. 
To fix a bug, a developer first analyzes the bug report looking for hints about bug location and then explores the source code to identify a few potential bug-related locations for which undesired behavior is later confirmed through debugging. To aid the developer in this process, and reduce time consuming project browsing, researchers have proposed many approaches to link a bug report to a specific section of the project code using automated Information Retrieval (IR) techniques~\cite{kim_buglocator_2013,saha_bluir_2013,chaparro_observed_2017,bug_scout}. IR-based Bug Localization techniques create an index from the preprocesed source code files, and subsequently use the index to build an IR model representing the software. When a new bug report arrives it is preprocessed in the same manner as the source code files, and it serves as a query to the IR model, which retrieves the most relevant program elements.


With modern software trending towards continuous deployment and relying on dynamic repositories with extremely high code churn and number of contributing developers \cite{potvin2016google,bhagwan2018orca,pradel2020}, new challenges for IR-based Bug Localization come to the forefront. First, the vast majority of prior work on IR-based Bug Localization has relied on files or methods taken from a snapshot of the source code, typically the most recent release of the software. However, as the software changes and evolves rapidly, the index and the IR model built on the snapshot of the source code becomes quickly outdated which can lead to the loss of bug localization effectiveness. At the same time, re-creating the index and the model incurs significant cost even on the latest hardware~\cite{lee2018} and leads to increased latency when retrieving the results~\cite{rao2013incremental}. 
To introduce the dynamics of software evolution into bug localization techniques, researchers have recently investigated using changesets (or commits) as the main unit of data used to construct retrieval models~\cite{wen2016,corley2018,lin2021traceability,rao2013incremental}. The advantage of changeset-based techniques is that they allow for continuous updates to the bug localization model, such that the model is built incrementally (or online) as changesets arrive, and captures all information about the source code at any given point in time.
Online Bug Localization is a new perspective on bug localization that uses changesets to avoid costly retraining as the software changes. 

Secondly, given that the modern software is a result of efforts of multiple participants (e.g., developers, testers, end users) with varying level of knowledge and expertise, bug reports may include different types of information, which has been observed to have a strong influence on the performance of IR-based bug localization techniques~\cite{wang2015usefulness,rahman2018improving}.  Bug reports differ based on how they describe the software failure, e.g., when bug reports are written by expert developers they tend to include detailed information about the source of the bug, such as stack traces or even direct references to relevant source code artifacts~\cite{mills2020relationship}. Bug localization for such bug reports is most effective when it relies on simple token matching between the bug report text and source code.  Another subset of bug reports, likely created by end-users that are unfamiliar with the software design or source code, provide only high-level description of the observed faulty behavior~\cite{hooimeijer2007modeling}. For this second category of bug reports more sophisticated bug localization techniques that, e.g., mine revision histories or build higher-level representations of the source code, are required. In this paper, we describe an adaptive approach that is able to adjust to the type of content present in the bug report of interest.


To perform bug localization for rapidly evolving software and in the presence of diverse bug reports, this paper proposes \jitbl, a model for online bug localization based on a streaming variant of the Latent Dirichlet Allocation (LDA) topic modeling technique~\cite{blei2003latent}, which has been previously employed by Corely et al.~\cite{corley2018} for the purpose of feature location. To capture the evolution of software artifacts and bug reports over time, we use two parallel Online LDA models, one that tracks changes in the source code repo and another that tracks bugs reported in the issue tracker. 
The models 1) naturally represent development activity, including frequently changed and co-changed program elements; 2) effectively handle streaming data, including the ability to deemphasize older information; and 3) raise the level of abstraction into topics, which provide a higher-level structure for detecting related artifacts. 
Translating between the probabilistic topic spaces maintained by the two models is performed using a translation matrix, primarily constructed from the history of fixed bug reports and their corresponding changesets. \jitbl adapts its prediction to the content of a bug report by emphasizing one of the probabilistic topic spaces. More specifically, the more code-like tokens are present in the bug report, the more weight is given to the LDA tracking source code. Conversely, if the bug report consist primarily of natural language, \jitbl relies more on the LDA tracking bug reports and translation matrix for the prediction. 

The main contributions of this paper are:


\begin{itemize}
\item
Online, up-to-date model for bug localization that integrates adaptability to different types of bug report content.
\item
Dataset of high code churn open-source Java repositories with mapping of bug reports to their fixing changesets.
\item
Evaluation results using our dataset and an established benchmark of software projects with various development history lengths.  The model and our experiments are available for further research as part of our replication package.\footnote{https://github.com/aciborowska/jingo}
\end{itemize}

\jitbl is situated in the current state of the art in the following way. Wen et al.'s technique (Locus) was the first to build a bug localization model based on changesets, with considerable success~\cite{wen2016}, however, their model was not designed to be frequently updated. Corley et al. were the first to explore the concept of continually updated LDA model based on changesets~\cite{corley2018}. Compared to Corley et al., we introduce an online bug localization technique that is adaptable to the diversity of bug reports observed in modern software repositories~\cite{catolino2019,zhang2018}.

\section{Background}

The bug localization problem is often framed as an Information Retrieval (IR) problem:
given a text of a bug report (i.e. query) find the most appropriate program
elements (i.e. documents). Popular IR approaches like the vector space model, which model
a document using the frequency of each term in the document (term frequency) and the corpus (document frequency),
are often used due to their simplicity and effectiveness~\citep{kim_buglocator_2013,mills_enought_2018,wen2016}.
The vector space model requires that the exact words that appear in the query
(bug report) are matched in the document (program element).
 On the other hand, other
IR techniques, e.g., topic models or neural networks, build a higher-level representation of the corpus
content and use it as a reference point for comparing individual documents~\citep{bug_scout,wang2018bug,huo2017enhancing,xiao2018bug}. One such technique
is Latent Dirichlet Allocation (LDA), which
models a set of documents probabilistically via a set of topics that appear in those documents~\citep{lukins_lda_2010}.
Using LDA, the similarity between the bug report and program element is computed after
they are both projected into the topic space, i.e. a query and bug report match if they
express a similar distribution of topics. 
Another popular group of techniques are based on neural networks, however, compared to LDA, these techniques are less interpretable and have higher training computational cost.
Most IR techniques, including LDA, assume a static corpus
of documents that does not change significantly over time. 
On the other hand, Online Bug Localization recognize the evolution of documents (i.e. source code) over time and uses online model which can be efficiently update with a stream of newly arriving data (i.e. changesets, which capture modifications to the source code).

\noindent
\textbf{Online LDA. }
Latent Dirichlet Allocation (LDA) is a Bayesian probabilistic model designed to discover latent topics in large document corpora~\citep{blei2003latent}. Based on a collection of documents, LDA infers latent document-topic distribution and a corresponding topic-word (or topic-term) distribution, such that each document is represented as a mixture of topics, and each topic is described by a mixture of words (or terms). Inferring document-topic and topic-word distributions is equivalent to training LDA and produces an interpretable model that can be applied to previously unobserved documents to extract their lower-dimensional representation. 
LDA is configurable via two main hyperparameters, $\alpha$ and $\beta$, affecting the smoothness of document-topic and topic-word distributions. Hyperparameter $\alpha$ influences the document-topic distribution, such that increasing the $\alpha$ value causes a document to express more topics, whereas lowering $\alpha$ makes a document to be represented by fewer topics. Conversely, hyperparameter $\beta$ influences the topic-word distribution. Raising $\beta$ causes words to relate to multiple topics, while lowering $\beta$ produces more specific topics with words rarely repeating between topics.

Updating LDA with a new document requires retraining the entire model from the beginning. As this process introduces significant time delay and computational cost, ordinary LDA is inappropriate for use in streaming environments where new documents, such as bug reports or changesets, arrive continuously. Therefore, to model the dynamics of modern software development, we use a recently devised variant of LDA, Online LDA \citep{hoffman2010online}, which allows to incorporate new documents through a update procedure, without the need for complete model retraining. Online LDA introduces the hyperparameter $\kappa$, which influences how quickly older information in the document stream is forgotten. Increasing $\kappa$ causes the pace of forgetting to rise, thus older documents have smaller impact on the current topic distributions.

\noindent
\textbf{LDA on Changesets.}
IR models like LDA can be trained and applied on separate, but closely related datasets, which
reflect a similar vocabulary of terms. In this study, we follow the methodology proposed by Corley et al.~\citep{corley2018}, that is to train an LDA model
on changesets, and infer topic distributions of program elements (e.g., classes) from a snapshot of the
same repository to compare them to the topic distributions of bug reports in
order to perform bug localization. Such a model would retrieve program elements to the developer.
Alternatively, the LDA model can both train on changesets and retrieve changesets to developers.
Wen et al. argued that presenting the developer with a changeset may provide contextual
clues that are not available when retrieving program elements, however, most bug localization
techniques focus on retrieving program elements~\citep{wen2016}.
The model requires no change to alternate between these two configurations, as these
operations are performed on the already trained LDA model.

\section{Motivation}\label{sec:motivation}

In this section, we aim to motivate the need for the two key characteristics of our bug localization technique, 1) online bug localization that tracks rapidly-changing modern software, and 2) adaptability to the differences in bug reports observed in software repositories.

\subsection{Online Bug Localization}

As software changes are committed into a repository, the version control system computes
changesets (or diffs) between the original and changed software files.
Each changeset includes the content of lines that were
added, removed or modified across all of the project's files, and a commit message
that provides a summary of the change in natural language.
Commonly, changesets also include lines of context before and after
each change\footnote{In this paper, we use output of the {\tt git diff} command with default
settings, which uses the default (i.e. Myers) diff algorithm.}. The set of changesets for a particular project contains a
superset of the information contained in any one snapshot
of the source code repository. In other words, all information available in a
snapshot of the software can be reconstructed from the change history~\citep{alali_commit}.

In simple terms, Online Bug Localization allows for building an up-to-date model for retrieving program elements
based on the most current version of the source code, i.e., HEAD in each developer's local copy of
the software repo, rather than on prior versions. We deem that most of the time, the current version represents the
source code that the developer cares about, rather than past snapshots reflecting a recent release
of the software. 
To confirm the previous statement, we conducted a small survey on Reddit, asking the question {\em ``When working on finding the location of a user-reported bug, what version of the software do you use most of the time?''}. The question was posted in the r/AskProgrammers, r/SoftwareEngineering and r/AskProgramming subreddits. More than two thirds of the respondents, 21/31, indicated that they use {\em ``The most recent version of the software from version control''}, while 10/31 indicated they use {\em ``The version of the software in use by the reporting user''}. One of the respondents further stated that {\em ``I start with trying to reproduce the bug with the most recent version. If I can't, then I rollback to the version it was reported on to try to recreate it. I do quick and dirty until quick and dirty doesn't cut it.''} Another one noted that {\em ``I work with desktop applications, which are obfuscated and optimized, so debugging the released version can be difficult and time consuming''}. Yet another responded indicated a continuous deployment style of software engineering, i.e., {\em``I do web projects so I use the version in the repo, in my node\_modules folder. It doesn't matter whats latest, because my app is using the version is was compiled with.''}. Overall, we note that developers prefer to start bug localization process by using the current version of the source code.
\subsection{Advantages of Changesets}
A key design choice in IR-based techniques that are applied to source code is what constitutes a document: classes, methods, files, or changesets. The document choice has a strong
influence on, e.g., how topics in LDA model are formed, therefore utilizing changesets as documents brings several advantages inherent to this data type.

\noindent
{\bf Co-changed code.}  Changesets have the benefit of capturing co-change code, which is known to be an important predictor of software
maintenance activity~\citep{hassan2004predicting}. When leveraging changesets as a primary document dimension, co-changed code entities will often appear within the same document boundary. In the result, techniques utilizing co-occurring terms, like LDA, can recognize code entities that are often modified together leading to better IR performance.
To understand the relationship between co-changed program elements, we empirically examine whether frequently co-changed classes
are likely to be expressed by similar distributions of topics in a LDA model.
For this purpose, we divide the co-changed classes in 4 different popular
software projects (BookKeeper, OpenJPA, Pig and ZooKeeper) into three categories: 1) class
pairs that are co-changed more than or equal to $20\%$ of the time; 2) class pairs that are co-changed
less than $20\%$ of the time but more than or equal to $5\%$ of the time; and 3) class pairs
that are co-changed less than $5\%$. 
For instance, a class pair is considered co-changed more than or equal to $20\%$ of the time, when the classes share at least $20\%$ of commits in their respective modification histories.
For each of these three categories of class pairs, we compute the
cosine similarity of the topic distribution vectors of the class pairs using LDA model built from all of the changesets in each project. To limit the computational time and avoid corner cases, we only consider
the 100 most changed classes for each project. Higher cosine similarity among frequently co-changed classes
indicates that the co-change relationship is also reflected by the LDA model trained on changesets.
The results, shown in Table~\ref{tab:cochange}, demonstrate that the more often classes are co-changed, the more similar are their topic distributions, which indicates that the LDA model is able to inherently reflects the co-change relationship.

\newcolumntype{P}[1]{>{\centering\arraybackslash}p{#1}}

\begin{table}[]
\footnotesize
\centering
\caption{How the co-change relationship between classes is reflected in a LDA model built from changesets. The figure reports the average cosine similarity of different categories of co-changed classes, where the similarity is computed using the inferred LDA distribution of the text of each class.}
\begin{tabular}{p{0.9cm}P{3.0cm}P{3.0cm}P{3.0cm}}
\toprule
& \multicolumn{3}{c}{{\em Average cosine similarity of classes}} \\
\midrule
& co-changed $>=20\%$ of time & co-changed $<20\%$ and $>=5\%$ of time & co-changed $<5\%$ of time\\ \midrule
BookKeeper & 0.571 & 0.172 & 0.056 \\
OpenJPA & 0.418 & 0.152 & 0.059  \\
Pig & 0.591 & 0.198 & 0.062      \\
ZooKeeper & 0.251 & 0.132 & 0.107        \\
\bottomrule
\end{tabular}
\label{tab:cochange}
\end{table}

\noindent
{\bf Frequently changed code.} Another similar advantage is in frequently changed code, which is likely to appear in
many documents. Researchers and practitioners have observed that, over time, bug
disproportionally appear in code that has been recently and frequently
changed~\citep{memon2017taming,graves_fault_2000}. Therefore, observing recent
code changes, rather that only considering static source code snapshots, can have
significant value to bug localization techniques.

\subsection{Heterogeneity in Bug Reports}\label{sec:br-heterogeneity}
The content of bug reports can vary as some reports provide explicit localization hints through stack traces or code element names, while others contain only high-level textual description~\citep{rahman2018improving}. To illustrate the different types of bug reports and their properties, in Figure~\ref{fig:example-123} we show three exemplary bug reports we encountered when examining reports from the BookKeeper project. Each example contains a summary and description of the bug report and a list of fixed classes, sorted by to the number of changed lines.

\begin{figure}[th!]
	\centering
	\begin{subfigure}[t]{0.99\textwidth}
	\fbox{\begin{minipage}{\textwidth}
	{\scriptsize \textsf{
	  \vskip -0.1cm
		\textbf{Summary:} \textit{GarbageCollectorThread} exiting with ArrayIndexOutOfBoundsException.\\
	    \textbf{Description:} After completing compaction, \textit{GarbageCollectorThread} will do flush any outstanding offsets. When there is no offset present, its throwing following exception and exiting. \\
		  \texttt{\scriptsize{
				  [stack trace]\\
					at org.apache.bookkeeper.bookie.GarbageCollectorThread\\
					\indent\$CompactionScannerFactory.flush\\
					\indent(GarbageCollectorThread.java:175)\newline
					[stack trace continues]
		  }}
	}}
	\end{minipage}}
	\vskip 0.1cm
	{\parbox{\textwidth}
	{\scriptsize
	\textbf{Fixed:} \texttt{GarbageCollectorThread}}}
	\caption{BookKeeper-700; code references (CR) bug report.}
	\label{fig:example-1}
	\end{subfigure}

	\vspace{0.25cm}

	\begin{subfigure}[t]{0.99\textwidth}
	\fbox{\begin{minipage}{\textwidth}
	{\scriptsize \textsf{
   	\vskip -0.1cm
		\textbf{Summary:} \textit{AutoRecovery} should consider read only bookies.\\
		\textbf{Description:} \textit Autorecovery \textit{Auditor} should consider the readonly bookies as available bookies while publishing the under-replicated ledgers. Also \textit{AutoRecoveryDaemon} should shutdown if the local bookie is readonly.
	}}
	\end{minipage}}
	\vskip 0.1cm
	{\parbox{\textwidth}
	{\scriptsize
	\textbf{Fixed:} \texttt{Auditor, BookKeeperAdmin, BookieWatcher,
	BookiesListener, AuditorRecoveryMain, ReplicationWorker}}}
	\caption{BookKeeper-632; shared terms (ST) bug report.}
	\label{fig:example-2}
	\end{subfigure}

	\vspace{0.25cm}

	\begin{subfigure}[t]{0.99\textwidth}
	\fbox{\begin{minipage}{\textwidth}
	{\scriptsize \textsf{
		\vskip -0.1cm
		\textbf{Summary:} Fix for empty ledgers losing quorum.\\
		\textbf{Description:} If a ledger is open and empty, when a bookie in the ensemble crashes, no recovery will take place (because there's nothing to recover). This open, empty, unrepaired ledger can persist for a long time. If it loses another bookie, it can lose quorum. At this point it's impossible for the bookie to know that its an empty ledger, and the admin gets notified of missing data.
	}}
	\end{minipage}}
	\vskip 0.1cm
	{\parbox{\linewidth}
	{\scriptsize
	\textbf{Fixed:} \texttt{Auditor, ReplicationWorker, AutoRecoveryMain, BookKeeperAdmin, AuditorElector}}}
	\caption{BookKeeper-742; natural language (NL) bug report.}
	\label{fig:example-3}
	\end{subfigure}
	\caption{Three bug reports with different characteristics.}
	\label{fig:example-123}
\end{figure}

Figure \ref{fig:example-1} displays bug report \#600, which describes a situation causing the \textsf{GarbageCollectorThread} to throw an exception and provides the resulting stack trace. To fix this bug, a developer modified \textsf{waitEntrylogFlushed()} method in the  \textsf{GarbageCollectorThread} by adding a check for the offset size. This bug report provides a comprehensive source of information for the developer as it directly points to the class to be fixed, with additional information on finding the method to fix available in the  stack trace. We refer to this type of bug reports as \textit{code references (CR) bug reports}.

Bug report \#632, shown in Figure \ref{fig:example-2}, refers to the correct way of handling "readonly bookies when publishing ledgers". To fix the bug, the developer modified multiple classes, some of which were already mentioned in the report's text. We observed numerous common tokens shared between the bug report and modified classes: {\em readonly, available, bookie, publish}. Bug report \#632 highlights that bug reports, even if not providing explicit localization hints, still often contain common tokens that, when grouped into topics, reflect higher-level similarity between concepts presented in the bug report and in the source code, thus we refer to this type of bug reports as {\em shared terms (ST) bug reports}.

Figure \ref{fig:example-3} shows bug report \#742, describing a chain of faulty behaviors starting when "recovery is not performed for an open and empty ledger after bookie crashes". When fixing the bug, the developer modified multiple classes, none of which is mentioned in the report. This bug report poses the most challenging task for automated bug localization based on IR, since it provides only textual description with no code references and no unique terms that clearly map to source code locations. However, we observe that the set of modified classes for bug \#632 and \#742 is similar, indicating that there exists an upper layer abstraction that correlates the set of fixed classes to bug \#742. In fact, after further examination of the bug fixing history, we noticed that this set of classes was frequently modified together in bug reports that were addressing similar topics. As these bug reports do not provide any code references and are expressed in natural language, we refer to them as {\em natural language (NL) bug reports}.


Analyzing the examples of different types of bug reports presented above led us to the following two observations. Firstly, bug reports display different levels of details, requiring adopting different strategies to maximize performance of automated bug localization~\citep{rahman2018improving}. 
For instance, in the case of CR bug reports, it is sufficient to rely on matching code terms from a bug report to code tokens in a source code base. Conversely, ST bug reports can leverage topic similarity. However, neither of these approaches is able to address NL bug reports.
This leads to our second observation, namely that even when common tokens or topics are not present, correlated high-level concepts are still expressed by the bug report and source code~\citep{hooimeijer2007modeling} and can be identified by mining bug fixing history. In a results, a bug report can be matched to the most relevant code entities by examining bug fixing history to identify similar bug reports and their related code entities.


%

\section{\jitbl Model}\label{sec:model}

Performing online bug localization requires the ability to operate in an environment where incoming changes are immediately integrated into a model, which is able to detect both simple (i.e., near exact terms) and high-level similarities between the code and bug reports. In this paper we introduce \jitbl, a novel adaptable bug localization technique based on changesets. \jitbl separately models the streams of bug reports and changesets with individual Online LDA models~\citep{hoffman2010online}, obtaining two independent topic spaces, one for bug reports and one for changesets. To translate between the two topic spaces, \jitbl constructs a translation matrix based on the history of previously fixed bug reports, which captures a mapping between high-level concepts expressed in bug reports and their corresponding fixed program elements.

\begin{figure}[tb]
\centering
\includegraphics[width=0.45\linewidth]{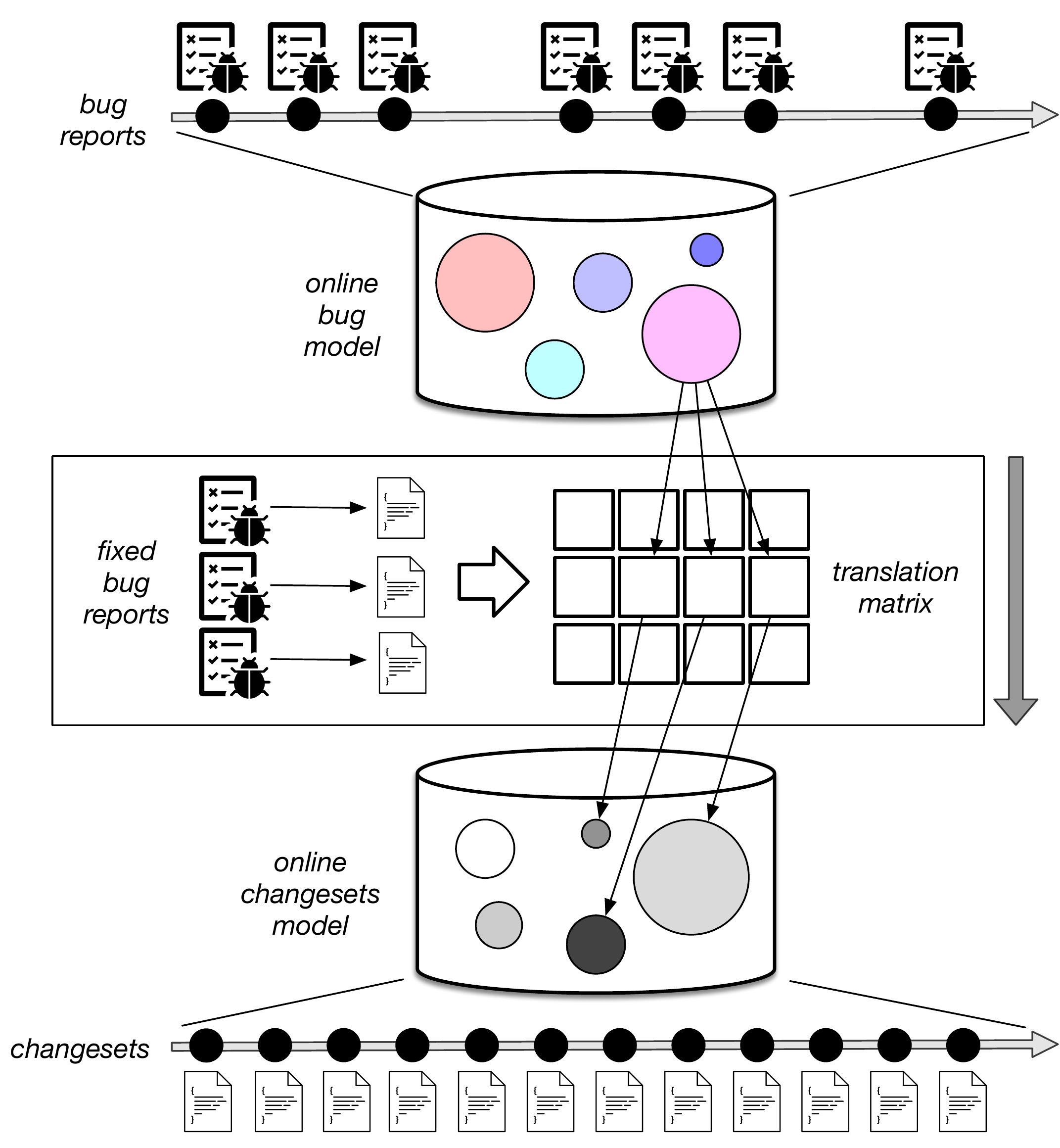}
\caption{The structure of the \jitbl\ model.}
\label{fig:bigpicture}
\end{figure}

The architecture of \jitbl, depicted in Figure~\ref{fig:bigpicture}, allows for dynamically adapting to the three different types of bug reports described in Section~\ref{sec:br-heterogeneity}. To this end, for a newly arriving bug report, \jitbl uses its changesets model and bug reports model to infer two topic distributions respectively. The first distribution is directed towards CR and ST bug reports that share code references or common concepts with the code base. The second distribution targets NL bug reports through the multiplication via the translation matrix.
The key idea behind the translation matrix is to utilize the bug fixing history in the project to capture the correlation between topics occurring in bug reports and in their relevant code entities. In other words, multiplying topic distribution of a bug report by the translation matrix results in a topic distribution of relevant code entities in the changeset topic space. 
Given that a bug report often include varying content, and hence, it is unlikely to be of only one type , we use a soft mechanism, based on the ratio of code tokens to all tokens in a bug report, to combine the two distributions. Finally, this combined distribution reflects the topics in the relevant code entities, and is used to select those elements in the code base.

\subsection{Structure of the \jitbl Model}

\jitbl is characterized by two parallel Online LDA topic models, one for changesets and the other for bug reports, and a matrix that translates from the bug reports to the changesets topic space, as shown in Figure~\ref{fig:bigpicture}.

\noindent
{\bf Changeset Model.} To build the changeset topic model, we use all changesets as they are committed into the source code repository. For every changeset, we use the output of {\tt git diff} command, which includes basic changeset information (e.g. commit SHA, author, date) and a list of changed code hunks, across all of the project's source code files, represented through added, modified or removed lines, each accompanied by 3 lines of context. We filter out the metadata and {\tt git diff} boilerplate formatting, such as e.g., \textsf{+++}, obtaining a set of file names and code modifications for each changeset\citep{amasaki2020}.
Following a recommendation of Eddy et al.~\citep{eddy2017}, we decided to give more weight to file names by repeating them 10 times to emphasize their importance to the Online LDA.
Finally, we follow the standard procedure to prepare source code for an IR model, including steps such as tokenization (using camel case and underscore), stemming using a Porter stemmer, and removal of standard programming language keywords, (e.g. \textsf{if}, \textsf{for}). In addition, we preserve the unsplit tokens into the corpus.  


\noindent
{\bf Bug Report Model.} We train the bug reports topic model with new bugs as they are reported in an issue tracking system, e.g., JIRA. For each report, we first retrieve its summary and description. The summary is commonly a single sentence, while the description provides more details about the bug. Finally, before updating the model, for each bug report we perform a preprocessing procedure common for natural language text, including tokenization, stemming and English stop word removal. Similarly as when building changeset corpora, we also preserve unsplit camel case tokens to ease locating relevant files when explicit code references are included in the bug report.


\noindent
{\bf Translation Matrix.} A translation matrix - $\matr{T}$ -  allows us to map from the bug report space to the changeset space, by simply multiplying a topic distribution inferred with the bug report model by the translation matrix, resulting in a projection into the changeset model's space of topics. The use of a translation matrix was inspired by TM-LDA, a model based on LDA that intends to predict the expected future topics for a stream of documents~\citep{wang2012tmlda}.

To create the $\matr{T}$ matrix we leverage previously fixed bug reports and their corresponding changesets.
The corpus of previously fixed bug reports provides an additional source of information that is leveraged
by many approaches to bug localization, e.g.~\citep{kim_buglocator_2013,rahman2018improving,ye2014learning,wang2020enhancing,alkhazi2020learning}.
However, the number of fixed bug reports depends on the size of the project and it is often very limited, thus using solely bug fixing history may not provide enough data to train the $\matr{T}$ matrix. To solve this cold-start problem, when building the $\matr{T}$ matrix we also include pairs of commit logs and changesets, since commit logs have been observed to contain substantial level of information describing in natural language the purpose or the functionality of the modified code~\citep{wen2016}.

We train the translation matrix using the following set of steps. First, for a set of fixed bug report - changeset pairs, and if necessary, commit log - changesets pairs, we infer a topic distribution for each fixed bug report using the bug report model and store it in matrix $\matr{B}$, where the rows of the matrix correspond to the distribution inferred for each of the fixed bug reports and the number of columns in $\matr{B}$ corresponds to the number of topics in the bug report LDA model. Second, an analogous procedure is performed for the matching changesets.
For these, the topic distribution is inferred by the changeset model and added to another matrix - $\matr{A}$, with rows containing topic distribution of changesets and column corresponding to the number of topic in the changeset model.
 In this way, each fixed bug report - changeset pair creates one corresponding row in matrices $\matr{B}$ and $\matr{A}$ respectively. Finally, training the translation matrix $\matr{T}$ reduces to solving the following equation.
\begin{align*}
\matr{T} =\argmin_{T}  \Vert \matr{B}\matr{T} - \matr{A} \Vert^2
\end{align*}
To solve the equation for the unknown $\matr{T}$ we perform least square minimization. Note that $\matr{T}$ is of size, number of topics in bug report model by number of topics in changeset model. Therefore, it requires \textit{at least} as many rows in $\matr{A}$ and $\matr{B}$, i.e., fixed bug reports, to compute. On the other hand, providing more than the required minimum amount of data to train the $\matr{T}$ matrix is desirable, as it is likely to increase the quality of mapping between topic spaces. We introduce a parameter specifying the minimum amount of data required to train the $\matr{T}$ matrix, $\omega$, expressed as a multiplying factor of the (maximum) number of topics required in the topic models. For instance, $\omega = 1.5$ and 50 LDA topics, would indicate that 75 fixed bug reports (or commit logs) are required to build the $\matr{T}$ matrix.

Once we determine the translation matrix, mapping between the bug reports to the changeset topic space is simple: we multiply the bug-related topic distribution for a new bug report by $\matr{T}$ matrix to get the equivalent distribution in changeset space. The computational cost of updating $\matr{T}$ over time is not large, and it is proportional to the number of fixed bug reports used. The cost can also be controlled by using a window based approach.


\subsection{Using \jitbl for Prediction}
\begin{figure}
\centering
\includegraphics[width=0.6\linewidth]{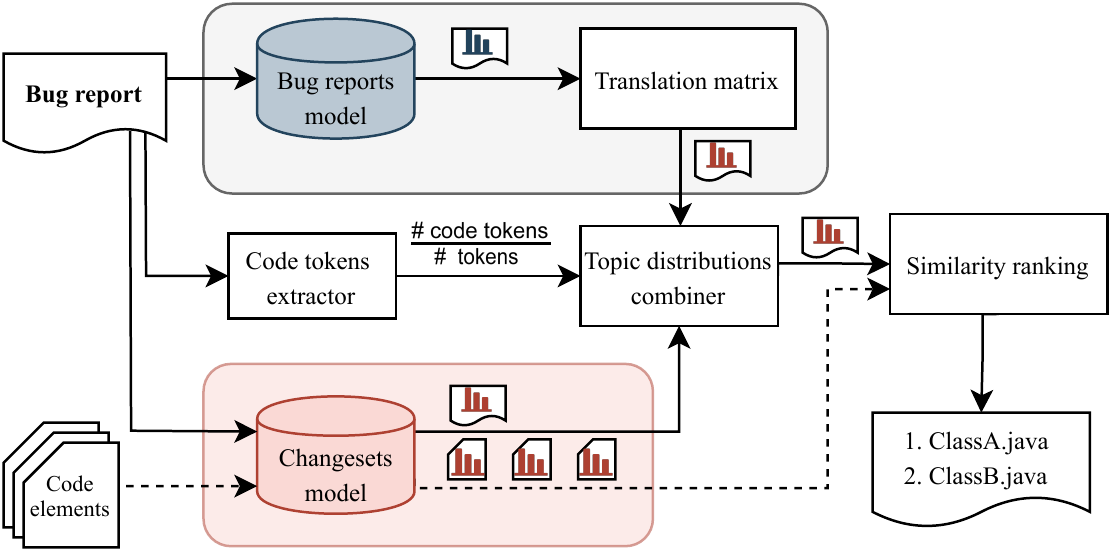}
\caption{Bug localization for a new bug report, where topic distributions in the bug report space (\protect\includegraphics[height=0.33cm]{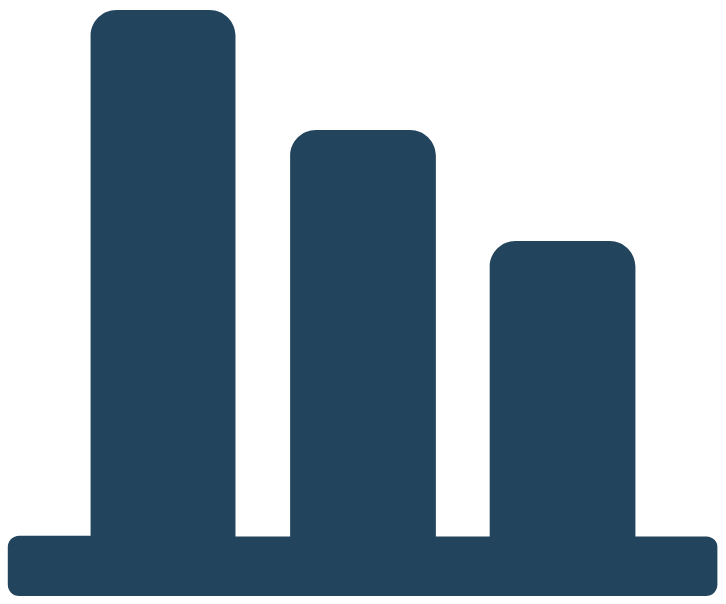}) are translated and combined with others in the changeset space (\protect\includegraphics[height=0.33cm]{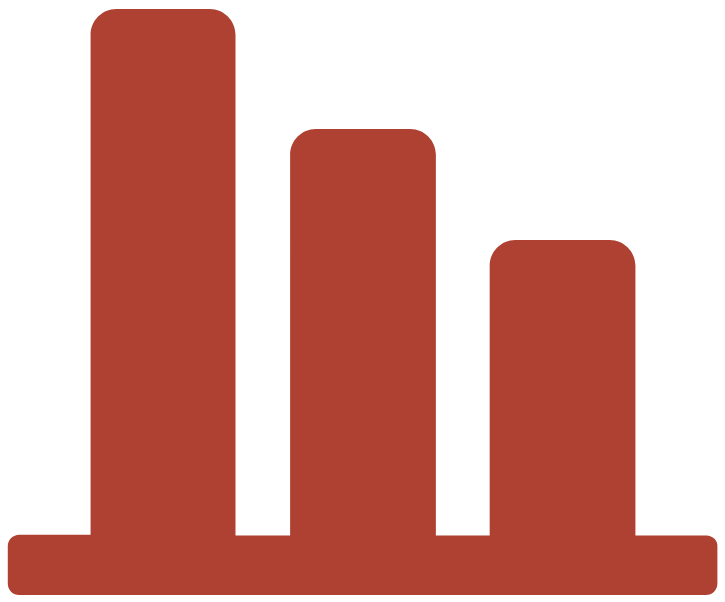}) to produce a combined topic distribution; cosine distance between the combined distribution and distributions for each of the code elements is used to produce the final ranking.}
\label{fig:bl-prediction}
\end{figure}

Using a trained \jitbl model, we follow the workflow in Figure~\ref{fig:bl-prediction} to perform bug localization for a newly arriving bug report.
To start, we preprocess the new bug report to construct a query, using the same procedure as when building the bug reports corpus. We use the query to infer two topic distributions, one using the changeset model -- {\em changeset-related topic distribution}, and another using the bug report model -- {\em bug-related topic distribution}. To map the {\em bug-related topic distribution} from the bug report model to the changeset model, we multiply it by the $\matr{T}$ matrix, obtaining a {\em co-occurrence topic distribution}. Note that, if the $\matr{T}$ matrix is not trained due to lack of data, this step can be skipped and the final prediction is then based solely on the {\em changeset-related topic distribution}.

Bug reports typically include various content and are likely to be related to the code base in more than one way via, e.g., code element names, shared tokens or bug fixing history. Hence, we decide to combine the {\em changeset-related topic distribution} and the {\em co-occurrence topic distribution} to better reflect bug report's characteristic.
To this end, we adopt a weighting strategy based on the number of code tokens that appear in the bug report. A token is considered to be a code token if it is in a camel case format or it corresponds to one of class names in the source code base.
We prioritize the importance of either of the distributions based on the intuition that the more code tokens are present in the bug report, the stronger is the similarity between the bug report and the source code base. If this is the case, we weigh the  {\em changeset-related topic distribution} more strongly. On the other hand, if code tokens are rare and the bug report predominantly consists of natural language text, then the {\em co-occurrence topic distribution} becomes more significant, as it leverages topics co-occurrences patterns that are not directly connected to bug's textual content. We use $\lambda$ to refer to the ratio of code tokens to the total number of tokens in a bug report. To account for the fact that natural language is more verbose in general and therefore the typical number of code tokens is much lower than the number of natural language tokens, we introduce an amplifying factor $\gamma$, which increases the importance of program element terms. We compute the {\em combined topic distribution} using the following equation:

$$
dist_{combined} = norm(dist_{changesets} * \lambda * \gamma + dist_{co-occurence} * (1-\lambda))
$$

Next, we use the changeset model to infer topic distributions for all documents in the most recent snapshot of the source code. As noted by researchers, bugs often pertain to small part of the code~\citep{ye2014learning,wong_brtracer_2014}, thus inferring at the granularity of large source code files, e.g., classes,  may negatively impact the performance of IR-based techniques. To solve this issue, we infer topic distributions of methods for each class and make a pairwise comparison against the combined topic distribution ($dist_{combined}$) of the query. Finally, each class is represented by a method that minimizes the cosine distance to the query and the classes are ranked according to increasing cosine distance to create a recommendation list.


\section{Evaluation setup}\label{sec:eval}

\subsection{Datasets: High Code Churn Dataset \& Bench4BL}

One of the key advantages of \jitbl is the ability to perform quick model updates as the new data arrives, making it suitable for repositories exhibiting high code churn. To gauge \jitbl\ performance for such projects, we curate a High Code Churn Repositories (HCC-Repo) dataset by selecting open-source projects hosted on GitHub.
To locate repositories of interest on GitHub, we perform a repository search query for all active Java repositories with at least one commit in 2019, and size of at least 10MB. Next, we sort the repositories in descending order by their code churn, calculated as the average number of commits per day, and select the top 10 repositories that use an explicit label to mark a reported issue as a bug (e.g., {\em bug}, {\em kind:bug}).
To build a goldset that connects a bug report to its fixing changesets and to retrieve a list of modified files, for each project we manually investigate 20 randomly selected bug reports to identify the project's convention for linking an issue to a changeset.
In general, we note that developers tend to use keywords, such as ``fixes``, ``closes`` or ``resolves``, or a project name followed by an issue number. Hence, to link a bug report to its fixing commit, we test commit messages and pull requests against two types of regex, \texttt{keyword\_\#XXX} and \texttt{project\_name-\#XXX}, where \texttt{XXX} denotes an issue number. In the case of identifying multiple changesets for a bug report, we follow Bench4BL's approach.
Our final dataset of high code churn repositories, HCC-Repo, contains 10 projects and the total of 874 bug reports.

To observe how \jitbl\ performs for repositories with more constrained commit traffic, we also use Bench4BL dataset~\cite{lee2018}, a collection of bug reports and corresponding lists of fixed source code files extracted from 51 open-source projects. 
The structure of the Bench4BL dataset allows for immediate use for evaluating release-based bug localization techniques, however it is not immediately suitable for an online approach. The key missing component in the dataset are the explicit connections between bugs and fixing changesets, which are necessary to perform evaluation in the online setting (i.e. timestamp-based). To this end, we adapted Bench4BL's code to retrieve the required data in a way that maintains complete consistency with the original dataset. More specifically, following Bench4BL's prior approach, to link a bug to a changeset we searched for an explicit mention of the bug identifier in the commit message. If more than one changeset was related to a bug, we selected the latest changeset, which is Bench4BL's existing assumption. Due to difficulties when finding clearly discernible links between bugs and changesets, we excluded 5 projects from Bench4BL dataset: JBMeta, ENTESB, ZXing, WLFY and SOCIALLI. Our final evaluation Bench4BL dataset is shown in Table~\ref{tab:dataset} and consists of 46 projects and 2,125 bug reports.

\newcolumntype{C}{>{\raggedleft\arraybackslash}m{1.7cm}}
\newcolumntype{R}{>{\raggedleft\arraybackslash}m{0.8cm}}
\newcolumntype{V}{>{\raggedleft\arraybackslash}m{1.7cm}}
\newcolumntype{B}{>{\raggedleft\arraybackslash}m{1.7cm}}
\newcolumntype{P}{>{\raggedright\arraybackslash}p{2.4cm}}

\begin{scriptsize}

\begin{longtable}{c|P|CVBB}
\caption{Projects used in the study. }
\label{tab:dataset}
\\ \toprule
\textbf{Group} & \textbf{Project} & \textbf{Commits} & \textbf{Evaluated version} & \textbf{Previously fixed bug} &  \textbf{Fixed bugs in version} \\ \midrule \midrule
\multicolumn{6}{l}{\textit{Hyperparameter tuning projects \citep{corley2018}}}                                                                                    \\ \midrule
\parbox[t]{2mm}{\multirow{4}{*}{\rotatebox[origin=c]{90}{\corleyshort}}}
& BookKeeper                        & 574          & 4.3.0                   & 223  & 102                            \\
& OpenJPA                           & 4,616         & 2.3.0                  & 1039 & 100                            \\
& Pig                               & 2,584               & 0.14.0                 & 1063 & 155                            \\
& ZooKeeper                         & 1,245        & 3.5.0                  & 368 & 235 \\
 \midrule
& Total & 9,019 & -- & 2,693 & 592
\\ \midrule \midrule
\multicolumn{6}{l}{\textit{Evaluation projects - HCC-Repo}} \\ \midrule
 \parbox[t]{1mm}{\multirow{10}{*}{\rotatebox[origin=c]{90}{HCC-Repo}}}
& DataHelix                  & 6,153              & --             & 107		  & 54                            \\
& eXist                          &  18,612                      & --             & 286         & 86                      \\
& Flink                           & 8,579             & --             & 424         &  100                        \\
& Hazelcast                  & 8,728                      & --             & 	425	       & 100                             \\
& Magarena                  & 24,883                       & --             & 	203	           &  72                            \\
& MegaMek                  & 18,899           & --             &  163         & 100 \\
& Micronaut                 & 9,469              & --             &  302         & 100 \\
& OpenDJ SDK            & 7,868               & --             &  358       & 100 \\
& ShardinSphere         & 27,014             & --             &  383         & 100 \\
& WooCommerce4A.  & 10,658             & --             &  102        & 62 \\
 \midrule
\multicolumn{6}{l}{\textit{Evaluation projects - BENCH4BL}}                                                                                               \\ \midrule
\parbox[t]{1mm}{\multirow{10}{*}{\rotatebox[origin=c]{90}{COMMONS}}}
& Codec                  & 1,387              & 1.5                    & 13		& 11                             \\
& Collections            & 2,837             & 4.0                    & 4		& 49                             \\
& Compress               & 1,033              & 1.4                    & 41		& 12                             \\
& Configuration          & 2,743             & 1.7                    & 66		& 31                             \\
& Crypto                 & 548              & 1.0.0                   & 1		& 8                              \\
& CSV                   & 1,085             & 1.3                    & 10		& 5                              \\
& IO                    & 1,850            & 2.0                    & 23 		& 25                             \\
& Lang                   & 5,231             & 3.5                    & 217 	& 40                             \\
& Math                   & 5,795             & 3.0                    & 74		& 39                             \\
& Weaver                 & 420              & 1.3                     & 0		& 2                              \\ \midrule
\parbox[t]{2mm}{\multirow{4}{*}{\rotatebox[origin=c]{90}{APACHE}}}
& Camel                           & 3,7986          & 2.15.0              & 1,298	     & 147                             \\[0.8ex]
& Hbase                           & 16,015            & 2.0.0                & 446		 & 418                            \\[0.8ex]
& Hive                              & 10,096           & 2.1.0                 & 547        & 221                             \\[0.7ex] \midrule \pagebreak \midrule
\parbox[t]{1mm}{\multirow{25}{*}{\rotatebox[origin=c]{90}{SPRING}}}
& AMQP                    & 1,254            & 1.5.0                  & 78		& 12                             \\
& Android                 & 504               & 1.0.0                   & 0		& 6                              \\
& Batch                     & 4,991            & 2.1.0                  & 460 	& 25                             \\
& Batch Admin          & 444              & 1.2.0                   & 4		& 5                              \\
& Data Commons     & 1,356            & 1.13.0                 & 140		& 20                             \\
& Data GemFire        & 1,123             & 1.4.0                  & 47		& 25                             \\
& Data JPA                & 875               & 1.11.0                  & 131		& 18                             \\
& Data MongoDB      & 1,879             & 1.5.0                  & 139		& 28                             \\
& Data Neo4j             & 461                & 4.0.0                   & 6		& 36                             \\
& Data Redis             & 1,421              & 1.8.0                  & 38		& 16                             \\
& Data Rest               & 883                & 2.6.0                   & 116		& 22                             \\
& Framework             & 14,635           & 3.0.0                 & 0 		& 13                             \\
& Hadoop                   & 1,473             & 2.1.0                  & 18		& 18                             \\
& LDAP                       & 979                & 1.3.0                   & 11		& 12                             \\
& Mobile                     & 324                & 1.1.0                   & 2		& 8                              \\
& Roo                         & 6,305             & 1.1.0                 & 201 	     & 134                            \\
& Security                  & 5,959             & 3.1.0                  &305		& 67                             \\
& Security OAuth       & 1,149             & 2.0.0                  & 112		& 9                              \\
& Shell                        & 275                & 1.1.0                   & 3 		& 2                              \\
& Social                      & 1,721              & 1.1.0                  & 11		& 6                              \\
& Social FB     & 1,265             & 1.1.0                  & 1		& 7                              \\
& Social Twitter         & 1,090             & 1.1.0                  & 2		& 4                              \\
& Webflow                 & 2,438             & 2.0.0                  & 24		& 24                             \\
& Web Service           & 2,042             & 1.5.0                  & 50 		& 23                             \\ \midrule

\parbox[t]{1mm}{\multirow{5}{*}{\rotatebox[origin=c]{90}{WILDFLY}}}
& Arquillian             & 606                                  & 1.1.0                & 0 		& 1                              \\
& Core                    & 11,379                              & 3.0.0                & 125		& 120                            \\
& Elytron                & 2,415                                & 1.1.0                 & 2		& 20                             \\
& Maven Plugin     & 481                                   & 1.1.0                  & 0		& 5                              \\
& Swarm                & 3,152                                & 2016.10            & 55		& 9                              \\ \midrule
\parbox[t]{1mm}{\multirow{4}{*}{\rotatebox[origin=c]{90}{ECLIPSE}}}
& org.aspectj                  & 7,721             & 286                 & --		     & 286                            \\
& platform.swt               & 24,585          & 98                    & --           & 89                             \\
& jdt.core                       & 22,459          & 94                    & --		     & 88                             \\
& pde.ui                          & 12,722          & 60                     & --		     & 57                             \\
 \midrule 

& Total                             & 358,526      & --     & 6,799        & 2,999  \\ \bottomrule
\end{longtable}
\end{scriptsize}

\subsection{Metrics}
To measure the effectiveness of our bug localization approach, we use three  metrics commonly used in previous bug localization studies \citep{lee2018,corley2018,wen2016,bug_scout}:

\begin{itemize}[nosep]
\item
{\em Mean Reciprocal Rank (MRR)} reports on the average reciprocal rank for the set of bug reports. Given a bug report and a ranking of code elements, the reciprocal rank is computed as the multiplicative inverse of the first rank among the fixed classes. Intuitively, MRR quantifies techniques performance when locating the first relevant code element.
\item
{\em Mean Average Precision (MAP)} considers ranks assigned to all relevant code elements for a bug report. This metric captures techniques abilities to recommend all code entities that are related to a bug.
\item
{\em Top@k} metric expresses the accuracy of bug localization technique, when considering the top {\em k} positions in the ranking of code elements potentially linked to the bug report. The value of Top@k metric is calculated as the percentage of bugs for which corresponding relevant buggy code entities are located in the top {\em k} elements of the ranking.
\end{itemize}

\subsection{Hyperparameter Optimization}
Several researchers have highlighted the importance of hyperparameter tuning of topic models for software engineering applications~\citep{agrawal2018wrong,treude2019predicting}, therefore prior to the evaluation, we optimize hyperparameters using a separate dataset released as part of Corley et al.'s study~\citep{corley2018}.
The key parameters to optimize are shown in Table~\ref{tab:params} and include the Online LDA priors for the two streaming topic models, and the set of parameters introduced by \jitbl.
First, we optimized the two Online LDA models independently using two metrics, perplexity and coherence~\citep{binkley2014understanding,koltcov2014latent}, and, second, we optimized the bug localization parameters based on the MRR metric.
%
To decide on parameters values to investigate for the changeset model, we followed results reported in previous research~\citep{corley2018}. 
In the case of the bug report model we used characteristics of the bug reports corpora, such as e.g., the number of documents and unique words, in relation to similar changeset corpora.
For the decay factor we used three values that correspond to the minimum, mean and maximum possible value for that parameter.
Note that, we did not optimize priors $\alpha$ and $\beta$ explicitly, relying instead on the automated estimation approach implemented in the {\em gensim} topic modeling library. We computed coherence and perplexity after training each topic model with 25\%, 50\% and 75\% of the respective corpus stream, in order to avoid bias towards longer streams (or larger corpora), averaging the values of the metrics to assess the model's performance.  Finally, we selected the top parameters for both topic models.

Next, we searched over the parameters related to \jitbl, namely $\omega$ and $\lambda$. The first parameter is a multiplying factor of the minimum number of fixed bug reports the model needs to observe before building the translation matrix. 
In the case of $\lambda$, we selected a set of values based on similar experiments conducted by Wen et al.~\citep{wen2016}. Values marked with bold in Table~\ref{tab:params} represent the set of final optimal values for all the parameters that we used during evaluation.

\begin{table}[h]
\centering
\caption{Hyperparameters and their corresponding values used during grid search; selected, optimal values are in bold.}
\scriptsize
\begin{tabular}{lll}
\toprule
\multicolumn{1}{l}{\textbf{Component}}                     & \multicolumn{1}{l}{\textbf{Parameter}} & \multicolumn{1}{l}{\textbf{Value}}            \\ \midrule
\multicolumn{1}{l}{\multirow{2}{*}{Changeset model}} & \# topics -- $k$                 & $\lbrace 75, \boldsymbol{100}, 150, 200 \rbrace $ \\
\multicolumn{1}{c}{}                              & decay factor -- $\kappa$          & $\lbrace 0.5, \boldsymbol{0.75}, 1.0 \rbrace $     \\ \midrule
\multirow{2}{*}{Bug report model}                 & \# topics -- $k$                 & $ \lbrace 10, 25, \boldsymbol{50}, 100 \rbrace $  \\
                                                  & decay factor -- $\kappa$         & $\lbrace 0.5, 0.75, \boldsymbol{1.0} \rbrace $     \\ \midrule
\multirow{2}{*}{Bug localization}                 & fixed bug reports factor -- $\omega$              & $\lbrace 1, \boldsymbol{1.5}, 2.0 \rbrace $       \\
                                                  & model combining factor -- $\gamma$                     & $\lbrace 1, 3, \boldsymbol{5}, 7 \rbrace $        \\
\bottomrule
\end{tabular}
\label{tab:params}
\end{table}

\subsection{Experimental Procedure}
To evaluate the performance of \jitbl we simulate the development history of a specific software project, continuously updating the model with bug reports and changesets as they arrive. 
More specifically, the bug report model is updated every time a new bug is reported, while the changeset model is updated when a new changeset is committed into the repository. 
Additionally, when a bug fixing changeset is observed, we also update the translation matrix with the fixed bug report and the fixing changeset.
Therefore, when evaluating for a specific newly arriving bug report, the changeset model contains all changesets that occurred before the time of \textit{the bug fixing commit}, while the bug report model includes all bug reports reported before the commit timestamp.


To evaluate the statistical significance of the difference in performance between \jitbl\ and the baseline, we compute the Wilcoxon signed-rank test with Holm correction and effect size using the Cliff's delta $\delta$. The values of $\delta$ ranges from +1 to -1, where -1 implies that all values in the first group are larger than values in the second group, and +1 represents the opposite situation. The effect size was intepreted using the following criteria: (1) small effect = $|\delta| > 0.147$; (2) medium effect = $|\delta| > 0.33$; and (3) large effect = $|\delta| > 0.474$ \citep{delta}. Note that in the case of small projects with number of bug reports equal or less than 10, we report the evaluation metric but did not conduct statistical testing.

\subsection{Research Questions}

\noindent\textbf{RQ1: How accurate is \jitbl in locating source code files relevant to a bug report?}
To answer RQ1, we use the proposed approach to identify buggy files in Bench4BL and high code churn datasets, and measure the effectiveness of \jitbl with respect to the previously defined metrics.
We compare the performance of \jitbl\ against the online technique proposed by Corley et al. \citep{corley2018}. Corley et al.'s technique has several similarities to \jitbl, as it uses a single Online LDA model trained on changesets to locate program elements relevant to a given bug report. The main difference is in that \jitbl models bug reports via a separate topic model and, through the usage translation matrix, incorporates information about previously fixed bug reports. Since Corley's approach is based solely on Online LDA trained with changesets, we set its hyperparameters to values identified to perform best for the changeset Online LDA part of our model (as described in Section 5.2). 
%

\noindent
\textbf{RQ2: What is \jitbl 's time overhead required to update the model?}
The key advantage of using an online model is the ability to update the model with new data once it arrives. However, a key question remains: how rapid is an update procedure when compared to a full model rebuild? To answer this question, we collect the execution logs for all studied projects and, based on the recorded timestamps, we compute the time required to {\em build} and to {\em update} the model. Specifically, \textit{build time} is the time required to build the model from scratch to the target version of the project we run the evaluation on, and \textit{update time} reflects the time needed to update the model with one changeset (i.e., a single commit).

\noindent
\textbf{RQ3: How does \jitbl compare to static (i.e., non online) bug localization techniques in terms of time overhead and accuracy?}
Most bug localization approaches proposed thus far use a static, snapshot-based model of the software. Although online models in general, and \jitbl in particular, focus on updating a model as the software changes, we still need to contrast its bug localization accuracy to state of the art static models. In this research question, we compare the accuracy of \jitbl in retrieving relevant results to state of the art static approaches based on the Vector Space Model (VSM). We also quantify the time overhead to rebuild the model for such techniques and contrast it to \jitbl.

\noindent
\textbf{RQ4: Can \jitbl adapt to different types of content in bug reports?}
Bug reports have diverse characteristics and can embody different level of details, that, when leveraged, can increase the effectiveness of a bug localization technique. The aim of RQ4 is to investigate how well the proposed model captures different types of bug reports. 
To this end, we randomly sampled a set of 322 bug reports from our corpus (95\% confidence level with a 5\% margin error to the target bug report population). The sample spanned 40 different projects. Subsequently, one of the authors manually categorized each of the bug reports into one of the three groups (i.e., CR, ST, and NL). We report MAP and MRR scores contrasted to Corley's~et~al.'s approach.

\newpage
\section{Results}

\newcolumntype{M}{>{\raggedright\arraybackslash}p{0.9cm}}
\newcolumntype{U}{>{\raggedright\arraybackslash}p{0.9cm}}
\newcolumntype{F}{>{\raggedright\arraybackslash}p{0.2cm}}
\newcolumntype{G}{>{\raggedright\arraybackslash}p{0.15cm}}
\newcolumntype{P}{>{\raggedright\arraybackslash}p{2.3cm}}
\renewcommand{\tabcolsep}{5pt}

\begin{scriptsize}
\begin{longtable}{P|MU|MU|MU|MU|MU}
\caption{Evaluation results for \jitbl\ compared to Corley et al.~ \citep{corley2018}, denoted as \textit{Cor.}. The per-project higher value of a metric is highlighted by light gray background (\colorbox{gray}{n.nnn}). Per-group of projects we used dark gray background (\colorbox{ggray}{n.nnn}) to highlight the higher value. Statistically significant increase in MRR and MAP values ($p$-value $<$ 0.05) is marked with bold type with a superscript indicating the effect size: $s$ -- small, $m$ -- medium, $l$ -- large. Projects marked with $\dagger$ had less than 10 bug reports, thus statistical testing was not conducted.}
\label{tab:results}

\\ \toprule
  & \multicolumn{2}{c|}{\textbf{MRR}} & \multicolumn{2}{c|}{\textbf{MAP}} & \multicolumn{2}{c|}{\textbf{Top@1}} & \multicolumn{2}{c|}{\textbf{Top@3}} & \multicolumn{2}{c}{\textbf{Top@5}} \\
 \textbf{Project} & \multicolumn{1}{r}{\textbf{\jitblshort}} & \multicolumn{1}{r|}{\textbf{\corleyshort}} & \multicolumn{1}{r}{\textbf{\jitblshort}} & \multicolumn{1}{r|}{\textbf{\corleyshort}} & \multicolumn{1}{r}{\textbf{\jitblshort}} & \multicolumn{1}{r|}{\textbf{\corleyshort}} & \multicolumn{1}{r}{\textbf{\jitblshort}} & \multicolumn{1}{r|}{\textbf{\corleyshort}} & \multicolumn{1}{r}{\textbf{\jitblshort}} & \multicolumn{1}{r}{\textbf{\corleyshort}} \\ \midrule
     \multicolumn{11}{c}{\textit{HCC-Repo}} \\ \midrule
 DataHelix & 0.150 &  \cellcolor{gray}0.196 & 0.109 &  \cellcolor{gray}0.121 & 0.094 &  \cellcolor{gray}0.151 & 0.208 &  \cellcolor{gray}0.264 & 0.283 &  \cellcolor{gray}0.321 \\
 eXist &  \cellcolor{gray}\textbf{0.123\textsuperscript{s}} & 0.100 & 0.049 &  \cellcolor{gray}\textbf{0.057\textsuperscript{s}} & 0.250 & 0.250 &  \cellcolor{gray}0.321 & 0.286 &  \cellcolor{gray}0.357 & 0.310 \\
 Flink & 0.269 &  \cellcolor{gray}0.308 & 0.178 &  \cellcolor{gray}0.195 & 0.172 &  \cellcolor{gray}0.222 & 0.273 &  \cellcolor{gray}0.323 & 0.364 & 0.364 \\
 Hazelcast &  \cellcolor{gray}0.386 & 0.324 &  \cellcolor{gray}0.235 & 0.219 &  \cellcolor{gray}0.280 & 0.204 &  \cellcolor{gray}0.440 & 0.398 &  \cellcolor{gray}0.550 & 0.459 \\
 Magarena &  \cellcolor{gray}0.111 & 0.104 &  \cellcolor{gray}0.085 & 0.073 & 0.028 &  \cellcolor{gray}0.042 &  \cellcolor{gray}0.141 & 0.127 &  \cellcolor{gray}0.183 & 0.127 \\
 MegaMek &  \cellcolor{gray}\textbf{0.236\textsuperscript{m}} & 0.106 &  \cellcolor{gray}\textbf{0.152\textsuperscript{m}} & 0.076 &  \cellcolor{gray}0.182 & 0.111 &  \cellcolor{gray}0.313 & 0.141 &  \cellcolor{gray}0.404 & 0.182 \\
 Micronaut &  \cellcolor{gray}\textbf{0.208\textsuperscript{s}} & 0.158 &  \cellcolor{gray}\textbf{0.148\textsuperscript{s}}& 0.120 &  \cellcolor{gray}0.131 & 0.091 &  \cellcolor{gray}0.222 & 0.162 &  \cellcolor{gray}0.253 & 0.232 \\
 OpenDJ SDK &  \cellcolor{gray}0.225 & 0.200 &  \cellcolor{gray}0.152 & 0.133 &  \cellcolor{gray}0.155 & 0.113 & 0.216 & 0.216 &  \cellcolor{gray}0.299 & 0.289 \\
 ShardingSphere &  \cellcolor{gray}\textbf{0.172\textsuperscript{s}} & 0.131 &  \cellcolor{gray}\textbf{0.128\textsuperscript{s}} & 0.101 &  \cellcolor{gray}0.234 & 0.202 & 0.287 & 0.287 &  \cellcolor{gray}0.351 & 0.330 \\
 WooComm.4A. &  \cellcolor{gray}\textbf{0.375\textsuperscript{m}} & 0.140 &  \cellcolor{gray}\textbf{0.194\textsuperscript{m}} & 0.093 &  \cellcolor{gray}0.295 & 0.082 &  \cellcolor{gray}0.443 & 0.197 &  \cellcolor{gray}0.541 & 0.279 \\
   \cmidrule{1-11}
 \textit{Group average} &  \cellcolor{ggray}\textbf{0.225\textsuperscript{s}} & 0.177 &  \cellcolor{ggray}\textbf{0.143\textsuperscript{s}} & 0.119 &  \cellcolor{ggray}0.182 & 0.147 &  \cellcolor{ggray}0.286 & 0.240 &  \cellcolor{ggray}0.358 & 0.289 \\
 \midrule
 \multicolumn{11}{c}{\textit{COMMONS}} \\ \midrule
 Codec         & 0.572                            & \cellcolor{gray}0.679 & 0.536                                        & \cellcolor{gray}0.621 & 0.364                           & \cellcolor{gray}0.545 & 0.818 & 0.818 & 0.818 & 0.818 \\
  Collections & \cellcolor{gray}0.629 & 0.591                           & \cellcolor{gray}0.490               & 0.478                          & \cellcolor{gray}0.490 & 0.429 & 0.694 & \cellcolor{gray}0.714 & \cellcolor{gray}0.796 & 0.776 \\
  Compress       & \cellcolor{gray}0.418 & 0.214 & \cellcolor{gray}0.255 & 0.133 & \cellcolor{gray}0.250 & 0.083 & \cellcolor{gray}0.500 & 0.167 & \cellcolor{gray}0.583 & 0.333 \\
  Configuration & \cellcolor{gray}\textbf{0.685\textsuperscript{s}} & 0.512 & \cellcolor{gray}\textbf{0.482\textsuperscript{s}} & 0.379 & \cellcolor{gray}0.548 & 0.387 & \cellcolor{gray}0.774 & 0.548 & \cellcolor{gray}0.903 & 0.677 \\
  Crypto\textsuperscript{$\dagger$} & \cellcolor{gray}0.654 & 0.453 & \cellcolor{gray}0.514 & 0.406 & \cellcolor{gray}0.571 & 0.286 & \cellcolor{gray}0.714 & 0.429 & \cellcolor{gray}0.714 & 0.571 \\
  CSV\textsuperscript{$\dagger$}  & 0.647 & \cellcolor{gray}0.717 & \cellcolor{gray}0.638 & 0.598 & 0.600 & 0.600 & 0.600 & \cellcolor{gray}0.800 & 0.600 & \cellcolor{gray}1.000 \\
  IO & \cellcolor{gray}0.694 & 0.627 & \cellcolor{gray}0.674 & 0.565 & \cellcolor{gray}0.560 & 0.520 & \cellcolor{gray}0.800 & 0.680 & \cellcolor{gray}0.840 & 0.720 \\
  Lang & \cellcolor{gray}\textbf{0.798\textsuperscript{l}} & 0.449 & \cellcolor{gray}\textbf{0.719\textsuperscript{l}} & 0.372 & \cellcolor{gray}0.700 & 0.300 & \cellcolor{gray}0.875 & 0.500 & \cellcolor{gray}0.925 & 0.650 \\
  Math & 0.584 & \cellcolor{gray}0.613 & \cellcolor{gray}0.474 & 0.453 & 0.462 & \cellcolor{gray}0.487 & \cellcolor{gray}0.641 & 0.615 & 0.744 & 0.744 \\
  Weaver\textsuperscript{$\dagger$}  & \cellcolor{gray}0.563 & 0.536 & \cellcolor{gray}0.549 & 0.490 & 0.500 & 0.500 & 0.500 & 0.500 & 0.500 & 0.500 \\
  \cmidrule{1-11}
  \textit{Group avg.} & \cellcolor{ggray}\textbf{\textit{0.624}\textsuperscript{s}} & \textit{0.539} & \cellcolor{ggray}\textbf{\textit{0.533}\textsuperscript{s}} & \textit{0.449} & \cellcolor{ggray}\textit{0.504} & \textit{0.414} & \cellcolor{ggray}\textit{0.692} & \textit{0.577} & \cellcolor{ggray}\textit{0.742} & \textit{0.679} \\
 \midrule 
  \multicolumn{11}{c}{\textit{APACHE}} \\ \midrule
 Camel & 0.258 & \cellcolor{gray}0.272 & 0.181 & \cellcolor{gray}0.192 & 0.163 & \cellcolor{gray}0.170 & 0.279 & \cellcolor{gray}0.293 & 0.374 & 0.374 \\
  HBase & \cellcolor{gray}0.393 & 0.355 & \cellcolor{gray}0.275 & 0.244 & \cellcolor{gray}0.278 & 0.246 & \cellcolor{gray}0.428 & 0.395 & \cellcolor{gray}0.524 & 0.462 \\
  Hive & \cellcolor{gray}\textbf{0.248\textsuperscript{s}} & 0.208 & \cellcolor{gray}\textbf{0.176\textsuperscript{s}} & 0.128 & \cellcolor{gray}0.137 & 0.133 & \cellcolor{gray}0.280 & 0.223 & \cellcolor{gray}0.365 & 0.303 \\
  \cmidrule{1-11}
  \textit{Group average} & \cellcolor{ggray}\textit{0.299} & \textit{0.278} & \cellcolor{ggray}\textbf{\textit{0.211}\textsuperscript{s}} & \textit{0.188} & \cellcolor{ggray}\textit{0.193} & \textit{0.183} & \cellcolor{ggray}\textit{0.329} & \textit{0.303} & \cellcolor{ggray}\textit{0.421} & \textit{0.380} \\ \midrule
  \multicolumn{11}{c}{\textit{SPRING}} \\ \midrule
 AMQP & \cellcolor{gray}\textbf{0.561\textsuperscript{m}} & 0.221 & \cellcolor{gray}\textbf{0.381\textsuperscript{m}} & 0.172 & \cellcolor{gray}0.417 & 0.083 & \cellcolor{gray}0.667 & 0.250 & \cellcolor{gray}0.750 & 0.333 \\
  Android\textsuperscript{$\dagger$}  & 0.239 & \cellcolor{gray}0.379 & \cellcolor{gray}0.243 & 0.195 & 0.167 & 0.167 & 0.167 & \cellcolor{gray}0.667 & 0.167 & \cellcolor{gray}0.667 \\
 Batch & 0.463 & \cellcolor{gray}0.621 & 0.288 & \cellcolor{gray}0.377 & 0.320 & \cellcolor{gray}0.480 & 0.480 & \cellcolor{gray}0.760 & 0.600 & \cellcolor{gray}0.800 \\
 Batch Admin\textsuperscript{$\dagger$}  & \cellcolor{gray}0.729 & 0.126 & \cellcolor{gray}0.542 & 0.118 & \cellcolor{gray}0.600 & 0.000 & \cellcolor{gray}0.800 & 0.200 & \cellcolor{gray}0.800 & 0.200 \\
 Data Commons & \cellcolor{gray}\textbf{0.569\textsuperscript{m}} & 0.343 & \cellcolor{gray}\textbf{0.483\textsuperscript{m}} & 0.285 & \cellcolor{gray}0.400 & 0.200 & \cellcolor{gray}0.750 & 0.350 & \cellcolor{gray}0.800 & 0.600 \\
 Data GemFire & \cellcolor{gray}0.645 & 0.532 & \cellcolor{gray}0.357 & 0.315 & \cellcolor{gray}0.520 & 0.360 & \cellcolor{gray}0.680 & 0.640 & \cellcolor{gray}0.840 & 0.720 \\
 Data JPA & 0.403 & \cellcolor{gray}0.410 & 0.312 & \cellcolor{gray}0.328 & 0.278 & \cellcolor{gray}0.333 & \cellcolor{gray}0.500 & 0.444 & \cellcolor{gray}0.556 & 0.500 \\
  Data MongoDB & \cellcolor{gray}0.531 & 0.418 & \cellcolor{gray}0.367 & 0.303 & \cellcolor{gray}0.357 & 0.286 & \cellcolor{gray}0.643 & 0.500 & \cellcolor{gray}0.786 & 0.571 \\
  Data Neo4j & 0.242 & \cellcolor{gray}0.262 & \cellcolor{gray}0.179 & 0.148 & 0.056 & \cellcolor{gray}0.167 & \cellcolor{gray}0.389 & 0.222 & \cellcolor{gray}0.444 & 0.333 \\
  Data Redis & \cellcolor{gray}0.586 & 0.505 & \cellcolor{gray}0.393 & 0.271 & \cellcolor{gray}0.500 & 0.375 & 0.625 & 0.625 & 0.625 & \cellcolor{gray}0.688 \\
  Data REST & 0.449 & \cellcolor{gray}0.504 & \cellcolor{gray}0.362 & 0.314 & 0.273 & \cellcolor{gray}0.318 & 0.500 & \cellcolor{gray}0.636 & 0.773 & 0.773 \\
  Framework & \cellcolor{gray}0.080 & 0.048 & \cellcolor{gray}0.060 & 0.038 & 0.000 & 0.000 & \cellcolor{gray}0.077 & 0.000 & \cellcolor{gray}0.154 & 0.000 \\
  Hadoop & 0.394 & \cellcolor{gray}0.434 & 0.300 & \cellcolor{gray}0.308 & 0.222 & \cellcolor{gray}0.333 & 0.444 & 0.444 & \cellcolor{gray}0.667 & 0.444 \\
  LDAP & \cellcolor{gray}0.427 & 0.258 & \cellcolor{gray}0.351 & 0.163 & \cellcolor{gray}0.250 & 0.167 & \cellcolor{gray}0.500 & 0.333 & \cellcolor{gray}0.667 & 0.333 \\
  Mobile & \cellcolor{gray}0.668 & 0.543 & \cellcolor{gray}0.634 & 0.510 & \cellcolor{gray}0.500 & 0.375 & \cellcolor{gray}0.750 & 0.625 & \cellcolor{gray}0.875 & 0.750 \\
  Roo & \cellcolor{gray}\textbf{0.221\textsuperscript{s}} & 0.184 & \cellcolor{gray}\textbf{0.191\textsuperscript{s}} & 0.141 & \cellcolor{gray}0.119 & 0.104 & \cellcolor{gray}0.246 & 0.187 & \cellcolor{gray}0.306 & 0.246 \\
  Security & \cellcolor{gray}0.469 & 0.410 & \cellcolor{gray}\textbf{0.356\textsuperscript{s}} & 0.311 & \cellcolor{gray}0.328 & 0.313 & \cellcolor{gray}0.552 & 0.448 & \cellcolor{gray}0.627 & 0.507 \\
  Sec. OAuth\textsuperscript{$\dagger$}  & \cellcolor{gray}0.353 & 0.261 & \cellcolor{gray}0.291 & 0.256 & \cellcolor{gray}0.222 & 0.111 & \cellcolor{gray}0.444 & 0.222 & 0.444 & 0.444 \\
  Shell\textsuperscript{$\dagger$}  & 0.556 & \cellcolor{gray}0.667 & 0.436 & \cellcolor{gray}0.440 & 0.500 & 0.500 & 0.500 & \cellcolor{gray}1.000 & 0.500 & \cellcolor{gray}1.000 \\
  Social\textsuperscript{$\dagger$}  & 0.554 & \cellcolor{gray}0.571 & \cellcolor{gray}0.483 & 0.474 & 0.500 & 0.500 & 0.500 & 0.500 & 0.667 & 0.667 \\
  Social FB\textsuperscript{$\dagger$}  & 0.545 & \cellcolor{gray}0.694 & 0.467 & \cellcolor{gray}0.592 & 0.429 & \cellcolor{gray}0.571 & 0.571 & \cellcolor{gray}0.714 & 0.571 & \cellcolor{gray}0.857 \\
  Social Twitter\textsuperscript{$\dagger$}  & 0.204 & \cellcolor{gray}0.773 & 0.235 & \cellcolor{gray}0.511 & 0.000 & \cellcolor{gray}0.750 & 0.250 & \cellcolor{gray}0.750 & 0.250 & \cellcolor{gray}0.750 \\
  Webflow & 0.282 & \cellcolor{gray}0.453 & 0.185 & \cellcolor{gray}\textbf{0.338\textsuperscript{s}} & 0.125 & \cellcolor{gray}0.333 & 0.375 & \cellcolor{gray}0.500 & 0.500 & \cellcolor{gray}0.542 \\
  Web Service & \cellcolor{gray}0.515 & 0.330 & \cellcolor{gray}0.379 & 0.237 & \cellcolor{gray}0.348 & 0.217 & \cellcolor{gray}0.609 & 0.348 & \cellcolor{gray}0.696 & 0.478 \\
 \cmidrule{1-11}
  \textit{Group average} & \cellcolor{ggray}\textbf{\textit{0.445}\textsuperscript{s}} & \textit{0.414} & \cellcolor{ggray}\textbf{\textit{0.345}\textsuperscript{s}} & \textit{0.298} & \cellcolor{ggray}\textit{0.310} & \textit{0.294} & \cellcolor{ggray}\textit{0.501} & \textit{0.474} & \cellcolor{ggray}\textit{0.586} & \textit{0.550} \\
 \midrule  \pagebreak \midrule
  \multicolumn{11}{c}{\textit{WILDFLY}} \\ \midrule
 Arquillian\textsuperscript{$\dagger$}  & \cellcolor{gray}1.000 & 0.200 & \cellcolor{gray}0.538 & 0.132 & \cellcolor{gray}1.000 & 0.000 & \cellcolor{gray}1.000 & 0.000 & 1.000 & 1.000 \\
  Core & 0.183 & \cellcolor{gray}0.208 & 0.139 & \cellcolor{gray}0.142 & 0.075 & \cellcolor{gray}0.133 & \cellcolor{gray}0.250 & 0.208 & \cellcolor{gray}0.308 & 0.267 \\
  Elytron & \cellcolor{gray}0.381 & 0.281 & \cellcolor{gray}0.355 & 0.262 & \cellcolor{gray}0.250 & 0.150 & \cellcolor{gray}0.400 & 0.350 & \cellcolor{gray}0.600 & 0.350 \\
  Maven Plugin\textsuperscript{$\dagger$}  & \cellcolor{gray}0.510 & 0.451 & \cellcolor{gray}0.459 & 0.357 & \cellcolor{gray}0.400 & 0.200 & 0.600 & 0.600 & 0.600 & \cellcolor{gray}0.800 \\
  Swarm\textsuperscript{$\dagger$} & \cellcolor{gray}0.341 & 0.203 & \cellcolor{gray}0.318 & 0.164 & \cellcolor{gray}0.222 & 0.111 & \cellcolor{gray}0.444 & 0.222 & \cellcolor{gray}0.556 & 0.333 \\
  \cmidrule{1-11}
  \textit{Group average} & \cellcolor{ggray}\textit{0.483} & \textit{0.269} & \cellcolor{ggray}\textit{0.362} & \textit{0.211} & \cellcolor{ggray}\textit{0.389} & \textit{0.119} & \cellcolor{ggray}\textit{0.539} & \textit{0.276} & \cellcolor{ggray}\textit{0.613} & \textit{0.550} \\
  \midrule
    \multicolumn{11}{c}{\textit{ECLIPSE}} \\ \midrule
  jdt.core & \cellcolor{gray}\textbf{0.258\textsuperscript{l}} & 0.066 & \cellcolor{gray}\textbf{0.124\textsuperscript{l}} & 0.019 & \cellcolor{gray}0.148 & 0.034 & \cellcolor{gray}0.295 & 0.068 & \cellcolor{gray}0.375 & 0.091 \\
  pde.ui & \cellcolor{gray}\textbf{0.192\textsuperscript{s}} & 0.149 & \cellcolor{gray}\textbf{0.130\textsuperscript{s}} & 0.101 & 0.105 & 0.105 & \cellcolor{gray}0.193 & 0.105 & \cellcolor{gray}0.316 & 0.193 \\
  platform.swt & \cellcolor{gray}\textbf{0.254\textsuperscript{m}} & 0.090 & \cellcolor{gray}\textbf{0.210\textsuperscript{m}} & 0.065 & \cellcolor{gray}0.135 & 0.034 & \cellcolor{gray}0.315 & 0.079 & \cellcolor{gray}0.393 & 0.112 \\
  org.aspectj & 0.041 & \cellcolor{gray}\textbf{0.113\textsuperscript{s}} & 0.022 & \cellcolor{gray}\textbf{0.041\textsuperscript{s}} & 0.021 & \cellcolor{gray}0.073 & 0.031 & \cellcolor{gray}0.119 & 0.042 & \cellcolor{gray}0.143 \\
  \cmidrule{1-11}
  \textit{Group average} & \cellcolor{ggray}\textbf{\textit{0.186}\textsuperscript{s}} & \textit{0.105} & \cellcolor{ggray}\textbf{\textit{0.122}\textsuperscript{s}} & \textit{0.056} & \cellcolor{ggray}\textit{0.102} & \textit{0.062} & \cellcolor{ggray}\textit{0.209} & \textit{0.093} & \cellcolor{ggray}\textit{0.282} & \textit{0.135} \\ 
  
 \midrule \midrule
  \textbf{Total} & \cellcolor{ggray}\textbf{0.415\textsuperscript{s}} & 0.352 & \cellcolor{ggray}\textbf{0.321\textsuperscript{s}} & 0.262 & \cellcolor{ggray}0.308 & 0.251 & \cellcolor{ggray}0.470 & 0.396 & \cellcolor{ggray}0.545 & 0.488 \\
 \bottomrule

\end{longtable}
\end{scriptsize}

\subsection{RQ1: How accurate is \jitbl in locating source code files relevant to a bug report?}

Table~\ref{tab:results} shows the performance of \jitbl alongside Corley et al.\citep{corley2018} baseline with respect to 5 metrics: MRR, MAP, Top@1, Top@3, and Top@5. Statistically significant improvements in \jitbl are marked with bold type with effect size (small, medium, large) noted in the superscript.
Overall, \jitbl\ demonstrates higher performance in locating buggy files across all of the above metrics with statistically significant increase of 6.3\% and 5.9\% for MRR and MAP respectively ($p$-value $<0.05$). At a finer scale, we observe an improvement in the average results across all groups of software projects, with statistically significant difference in MRR and MAP values for the COMMONS, SPRING, ECLIPSE and HCC-Repo groups and MAP value for the APACHE group.

Specifically, \jitbl\ achieves better performance in terms of MRR values for 37 out of 56 projects with a statistically significant improvement in 14 out of 43 projects that has more than 10 bug reports.
In the ECLIPSE group, MRR results for all but one project were improved with statistical significance, increasing MRR score obtained by the baseline by 19.2\%, 4.3\% and 16.4\% for {\em jdt.core}, {\em pde.ui} and {\em platform.swt}. In the case of {\em org.aspectj}, the baseline outperformed \jitbl\ by 7.2\%. We note the highest variance of increase and decrease of MRR values for the SPRING group, with \jitbl outperforming the baseline in 13 out of 24 projects. Finally, in the HCC-Repo group, \jitbl improved MRR scores for 8 out of 10 projects, including a statistically significant improvement observed for 5 projects.

Improvement in MAP values is achieved by \jitbl for 42 out of 56 projects with statistically significant difference noted for 14 projects. Similarly as for the MRR metric, the highest MAP increase of 8.4\% is observed for the COMMONS group with \jitbl\ outperforming the baseline for all but one project. We also observe the ECLIPSE projects achieved significantly higher results when using \jitbl\ with a small effect size. For the SPRING group, the proposed approach improves MAP results in 17 out of 24 projects. The improvement is statistically significant for 2 projects with medium ({\em AMQP}, {\em Data Commons}) and 2 with small ({\em Roo}, {\em Security}) effect sizes. In the HCC-Repo group, we note that \jitbl outperforms baseline for 7 out of 10 projects with a significant improvement in 4 projects.

On average, \jitbl\ outperforms the baseline for Top@1, Top@3 and Top@5 recommendation by 5.7\%, 7.4\% and 5.7\% respectively. For 4 project groups, APACHE, SPRING, ECLIPSE and HCC-Repo, we observe that as we consider longer recommendation list the difference between the techniques is growing. However, for the COMMONS and WILDFLY groups, the difference between \jitbl\ and Corley et al.'s approach reduces at Top@5 and Top@3 respectively.

\begin{figure}[t!]
    \centering
    \begin{subfigure}[t]{0.32\columnwidth}
        \centering
        \includegraphics[scale=0.68]{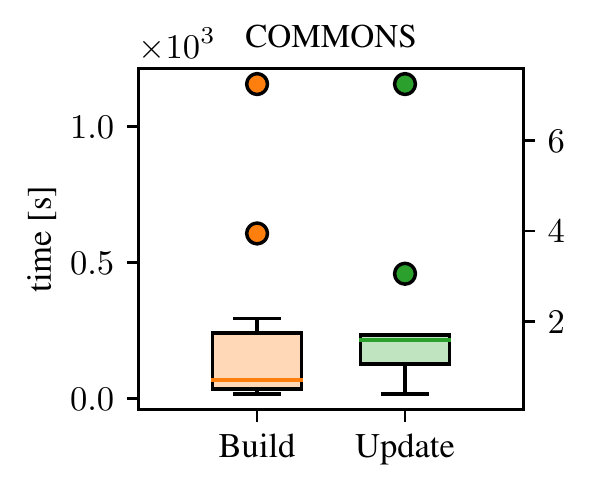}
    \end{subfigure}
    \begin{subfigure}[t]{0.32\columnwidth}
        \centering
        \includegraphics[scale=0.68]{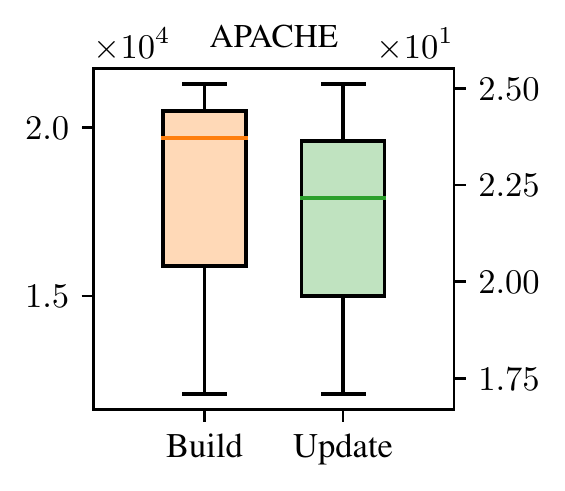}
    \end{subfigure}
    \begin{subfigure}[t]{0.32\columnwidth}
        \centering
        \includegraphics[scale=0.68]{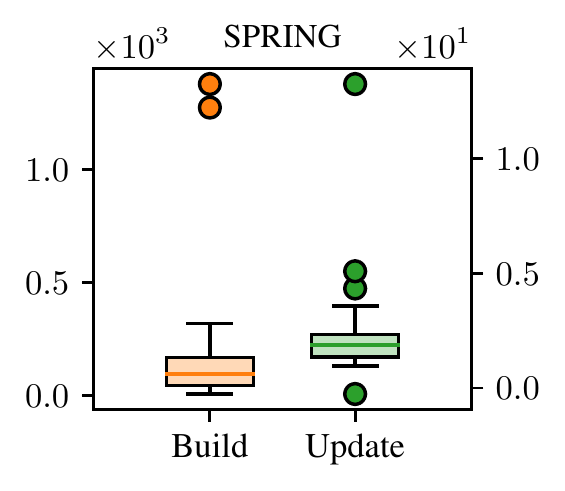}
    \end{subfigure}%

     \begin{subfigure}[t]{0.32\columnwidth}
        \centering
        \includegraphics[scale=0.68]{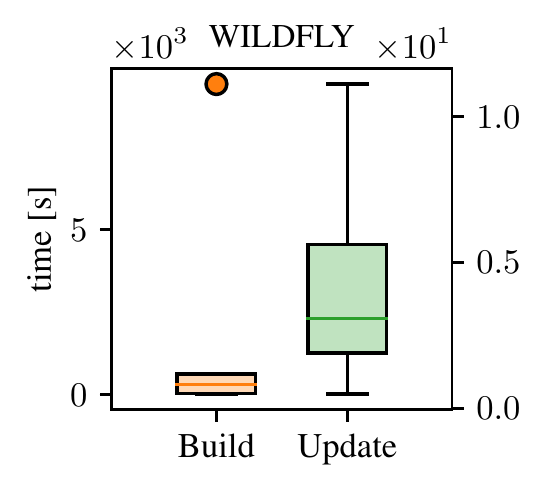}
    \end{subfigure}
    \begin{subfigure}[t]{0.32\columnwidth}
        \centering
        \includegraphics[scale=0.68]{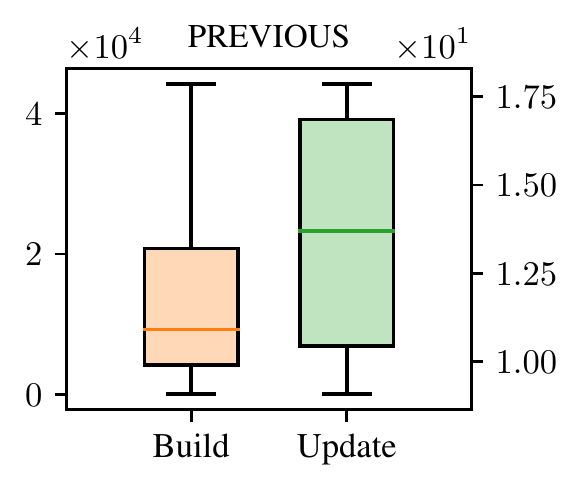}
    \end{subfigure}
    \begin{subfigure}[t]{0.32\columnwidth}
        \centering
        \includegraphics[scale=0.68]{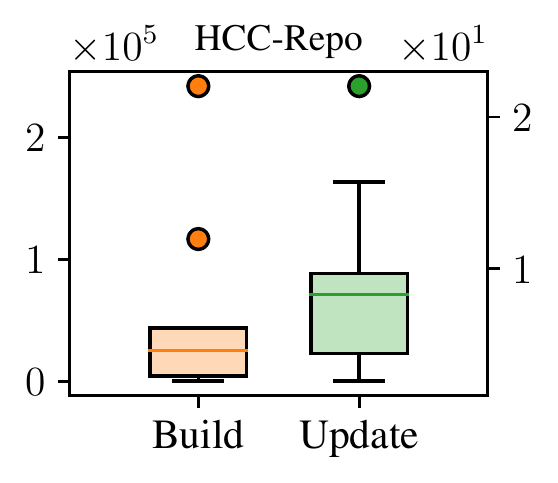}
    \end{subfigure}
    \caption{Average time in seconds required to \textit{build} and to \textit{update} \jitbl.  Note that \textit{bulid} and \textit{update} time are illustrated with two independent $y$ axes.}
    \label{fig:perfromance}
 \end{figure}

\subsection{RQ2: What is \jitbl 's time overhead required to update the model?}
Given the fact that \jitbl is an online model, it is updated with each newly arriving data instance, keeping the model up-to-date with minimal time overhead.
To investigate the performance benefit of using an online model over a model that requires re-building, in Figure~\ref{fig:perfromance} we show the time required to build and to update the model, averaged per each group of projects.
Across all groups, we observe that the update time is significantly lower than build time, with the average speedup of 100 for the groups with smaller projects (SPRING and COMMONS), and up to about 1000 in the case of large projects (APACHE, ECLIPSE and HCC-Repo). This indicates that with the growing size of a repository, the cost of re-training the model becomes even more prohibitive. As an example, consider the results obtained for APACHE projects, with the average build time close to $20,000s = 5.5h$ and update time of about $22.5s$. With new changesets being committed to a repository multiple times during a day, a model that relies only on re-building is outdated every couple of hours and consumes computational resources for a significant amount of time. On the other hand, utilizing an online model with an update procedure significantly reduces the time overhead, hence allowing to incorporate new information as it arrives.

\subsection{RQ3: How does \jitbl compare (in terms of time overhead and accuracy) to static (i.e., non online) bug localization?}

In order to compare \jitbl 's bug localization accuracy to that of the state of the art static models, we select two recent techniques based on the Vector Space Model (VSM), BLiA~\citep{youm2015bug} and BRTracer~\citep{wong_brtracer_2014}. We used source code for both of these techniques that was shared as part of Bench4BL~\citep{lee2018}. When considering all types of bug reports, simpler models like VSM have been reported to outperform LDA on bug localization~\citep{rao2011retrieval}. Table~\ref{tab:vs_static} shows the average MAP and MRR for \jitbl, BLiA and BRTracer on the Bench4BL set of repositories; we compute the average per bug report in order to account for some projects having more instances than others. Alongside the accuracy measures, Table~\ref{tab:vs_static} shows the average time overhead to construct the model and to update it with a single changeset, also averaged across all the projects in our dataset. 
BLiA and BRTracer do not have an update mechanism so the update time can be assumed to be equivalent to (re-)building the model.

BLiA and BRTracer outperforms \jitbl in retrieval accuracy with the improvement in MRR of 4.8\% and 14.8\% respectively. On the other hand, \jitbl offers significantly more time-efficient update procedure that is about 20 times faster than performing a full model rebuild for the VSM-based baselines.
Although \jitbl needs more time to initially construct the model when compared to  BLiA and BRTracer, note that  the build time depends on the number of documents. 
While BLiA and BRTracer use source code files (e.g., java class files),  \jitbl leverages changesets which are significantly more numerous compared to the number of classes, hence it is expected that \jitbl requires more time to complete the initial build.

\newcolumntype{K}{>{\raggedleft\arraybackslash}p{1.7cm}}
\begin{table}[t]
\begin{scriptsize}
\begin{tabularx}{\textwidth}{P|K|K|K|K}
\caption{Comparison between \jitbl and two VSM techniques based on average (per bug report) accuracy and time overhead measures.}
\label{tab:vs_static}
\\ \toprule 
& \multicolumn{2}{c|}{\textbf{Accuracy}} & \multicolumn{2}{c}{\textbf{Time overhead}} \\
\textbf{Technique} & \multicolumn{1}{c|}{\textbf{MRR}} & \multicolumn{1}{c|}{\textbf{MAP}} & \multicolumn{1}{c|}{\textbf{Build Time [s]}} & \multicolumn{1}{c}{\textbf{Update Time [s]}} \\
\midrule
\jitbl & 0.323 & 0.241 & 2964.955 & 4.786\\
BLiA & 0.371 & 0.314 & 101.315& 101.315$^{*}$ \\
BRTracer & 0.471 & 0.367 & 91.245 & 91.245$^{*}$ \\
\bottomrule
\end{tabularx}
\end{scriptsize}
\end{table}


\newpage
\subsection{RQ4: Can \jitbl adapt to different types of content in bug reports?}





One of the key goals of \jitbl\ is to adapt to different types of bug reports, namely Code References (CR), Shared Terms (ST) and Natural Language (NL) bug reports.
Table~\ref{tab:manual} contrasts the accuracy of \jitbl\ and Corley's~et~al.'s approach for the manually annotated set of bug reports.
In the case of CR bug reports, \jitbl\ improves upon Corley's et al.'s approach by 9\% in MRR, and 11\% in MAP, while for ST bug reports both techniques achieve comparable accuracy.
\jitbl\ demonstrates better accuracy for NL bug reports, improving MRR by 4.8\% and MAP by 5.2\% compared to Corley's et al.'s. Interestingly, comparing the drop in accuracy between ST and NL bug reports, we note that \jitbl\ lost about 3\% in each metric, while Corley's et al.'s performance decreased by about 9\% . This result indicates that \jitbl\ is in general more robust for NL bug reports, which we attribute to leveraging historical data via a translation matrix.

Overall, both techniques perform best with CR bug reports, followed with ST, and struggle the most with NL.  
Compared to Corley's et al.'s approach, \jitbl provides observable improvements in the CR and NL categories, but performs on par on ST bug reports.

\newcolumntype{Q}{>{\raggedright\arraybackslash}p{3.5cm}}

\begin{table}[t]
\begin{scriptsize}
\begin{tabularx}{\textwidth}{Q|M|MUU|MUU}
\caption{Performance on different bug report types from a manually annotated set ($N=322$).}
\label{tab:manual}
\\ \toprule
\multicolumn{2}{c|}{\textbf{Bug Report}} & \multicolumn{3}{c|}{\textbf{MRR}} & \multicolumn{3}{c}{\textbf{MAP}} \\
\textbf{Type} & \textbf{Num.} & \multicolumn{1}{r}{\textbf{\jitbl}} & \multicolumn{1}{r}{\textbf{\corleyshort}} & \multicolumn{1}{r|}{\textbf{diff}} & \multicolumn{1}{r}{\textbf{\jitbl}} & \multicolumn{1}{r}{\textbf{\corleyshort}}  & \multicolumn{1}{r}{\textbf{diff}} \\
\midrule
Code References (CR) & 165 & 0.455 & 0.365 & 0.090 & 0.376 & 0.266 & 0.110 \\
Shared Terms (ST) & 100 & 0.257  & 0.272 & \-0.015 & 0.180 & 0.177 & 0.003 \\
Natural Language (NL) & 57 & 0.226 &  0.178 & 0.048 & 0.143 & 0.091 & 0.052\\
\bottomrule
\end{tabularx}
\end{scriptsize}
\end{table}

\subsection{Discussion}

In this section, we detail a few salient observations resulting from evaluating our four research
questions.

The evaluation of RQ1 indicates that \jitbl presents an improvement over a prior technique that proposed
an updatable model of the software based on changesets.
More specifically, according to RQ4, the primary reason for this improvement is two fold, (1) \jitbl performs
better on bug reports with high number of code terms, including exact references to the methods and
classes of interest, (i.e., code references); and (2) \jitbl shows improvements on bug reports with
high level of abstraction that do not mention any of the related code references.
We believe the first reason is due to heuristics \jitbl uses, such as higher weighting program element names, while
the second reason is due to the two-level hierarchical architecture of \jitbl.
However, in the evaluation of RQ3, we observe that the accuracy of \jitbl is not at the level
of static (i.e., non online) approaches, in particular, those based on the vector space model.

\jitbl  achieves the goal of performing fast updates for newly arriving data, according to RQ2. The static VSM-based approaches we contrasted with in RQ3 are significantly faster to build than \jitbl, but are 20x slower in model updating, as these models are not designed to be updatable, hence they have to be rebuilt. 
More specifically, VSM typically uses tf-idf to represent each document as a vector of weights, such that weight $w$ of term $t$ in document $D$ is $w_t = tf(t) \times idf(t)$. While $tf$ (term frequency), which is computed as the number of times term $t$ occurs across all documents, can be easily updated as new documents are added by modifying term counts, the same cannot be said for $idf$ (inverse document frequency). The $idf$ values depend on the number of documents in the corpus and number of times terms occur in those documents, so $idf$ values need to be re-computed as documents are added. While it is certainly the case that this can be done periodically, the VSM model does not provide for an online approach to do so.

According to our evaluation of RQ4, the most abstract (Natural Language category) bug reports still perform poorly in absolute terms, i.e., MRR=0.226 and MAP=0.143, producing weaker results than the two other categories of bug reports we identified, despite the heavy emphasis of \jitbl on capturing abstract semantics via its two-layer architecture. Clearly, more work is needed to improve how bug localization techniques perform for this category of bug reports. Future research efforts should identify such bug reports explicitly as they are significantly fewer than the other categories (roughly 1/3 or 1/2 of the other categories in our randomly collected sample), while localizing them arguably provides the greatest value to end users.

\subsection{Threats to Validity}\label{sec:threats}
The results of the study presented in this paper suffer from several threats to their validity. A key threat to the internal validity of our study are the specific parameter
choices we used to build our model. Probabilistic models like ours are particularly sensitive to
such parameters~\citep{agrawal2018wrong}. While, to mitigate this threat we employed extensive hyperparameter
optimization using a separate dataset, it is clear that this threat can still be impacting our study.

Leveraging changesets for bug localization pose another threat due to possible noise that can be introduced by tangled, split, or
refactoring  changesets~\citep{mcintosh2011build,herzig2013tangled}.
However, as long as such noisy changesets are in the minority relative to ones that reflect semantically
related modifications, probabilistic techniques like LDA are likely to still produce a reasonable representation that can model how source code evolves over time~\citep{gelman2014bayesian}.

Another threat is in potential biases affecting our evaluation datasets, such as incorrect ground truth, and misclassified or already localized bug reports~\citep{kochhar2014}. 
The first two biases have a potential to negatively affect the performance of \jitbl as they introduce noise in the translation matrix, while the last bias can spuriously increase the results by having localization hints present.
To mitigate the first threat, we followed experimental procedures used by other
researchers, aiming in most cases to err on the side of caution by adopting choices that
produce low false positives when identifying source code files related to a bug~\citep{lee2018}. To mitigate the risk of using issues misclassified as bug reports, before building HCC-Repo dataset one of the authors manually inspected each project to ensure  the quality of issue labeling, and identified labels referring to actual bugs. Although our efforts cannot completely remove those two biases, as observed by Kochhar et al.~\citep{kochhar2014} they are typically not significant, hence, considering the size of our dataset, they should not have a significant effect.
As for the already localized bug reports, we did not exclude them since one of our goals was to observe how  bug localization performance changes for different types of bug reports. However, to present a complete picture, we included results for a manually annotated subset of bug reports with and without localization hints (Table~\ref{tab:manual}).

A key threat to external validity 
is that we applied the bug localization technique only on a limited number of bugs, which primarily
reflect popular open source Java projects. A mitigating factor is the evaluation with the large number
of projects curated by the Bench4BL benchmark~\citep{lee2018}.  Additionally, this benchmark has also
been applied to prior bug localization studies.
Another threat to external validity is in the chosen evaluation metrics, which may not directly correspond to
user satisfaction with our bug localization technique~\citep{wang2015usefulness}, impacting the generalizability
and validity of the reported results. We mitigate this threat by evaluating our approach with high-quality
datasets and well-known metrics, which continue to
be used by academia and industry to measure the performance of IR techniques.

\section{Related Work}\label{sec:related}

Automatically retrieving a list of code elements based on a newly written bug report has generated significant interest among
researchers for several years. In this section, we first outline the most recent and transformative approaches to IR-based bug localization, highlighting techniques that are able to adapt to rapidly changing software repositories, followed by an overview of recent evaluation techniques for bug localization.

At their core, techniques for IR-based bug localization rely on a
similarity measure between a bug report and code elements in the source code
base, which can be computed based on a variety of models and using different
sources of information found in the bug report or the source code. For instance, BugLocator~\citep{kim_buglocator_2013}
combines two rankings, one produced by similarity between the bug report and code elements
using a revised Vector Space Model (rVSM) and another based on similarity of the bug report to prior
fixed bug reports. BLUiR~\citep{saha_bluir_2013} improves over BugLocator by using program structure to boost specific terms (e.g., class names), while AmaLgam~\citep{wang_amalgam_2014} creates an ensemble consisting of BugLocator, BLUiR and a defect predictor leveraging development history of a project. BRTracer~\citep{wong_brtracer_2014} innovates by parsing and prioritizing stack traces that may occur in bug reports. Similar to our LDA-based model for bug localization, BugScout proposes
a modification of LDA that correlates bug reports and code elements via shared topics~\citep{bug_scout}. HyLoc uses a deep neural network to build connections
between the text in bug reports and that in code elements~\citep{lam2015}, while Xiao et al. explores structural and semantic information to discover relationships between bug reports and souce code~\citep{xiao2018bug}. Recently, Huo et al. proposed a novel convolutional neural network to learn unified feature representation from natural and programming language that captures both lexical and program structure information~\citep{huo2016learning}. This work was extended later on by modeling the sequential nature of source code using LSTM~\citep{huo2017enhancing}. To address the lack of historical data, Zhu et al. proposed approach based on adversarial transfer learning to detect and transfer common characteristics between projects~\citep{zhu2020cooba}. While most of these techniques, as
ours, benefit from observing more fixed bug reports to correlate to code elements, some do not
have an alternative for prediction and are likely to perform very poorly on projects with short histories. All of these techniques are built on static source code entities and not on changesets.

Researchers recognize that bug reports are diverse, and their content differences
can strongly influence the effectiveness of a bug localization technique. At the same time, it has
been reported that bug reports usually contain all the necessary information for effective IRBL~\citep{mills_enought_2018}. In order to reduce the noise present in bug report and focus IRBL on the most relevant terms, Chaparro et al. present a query reformulation strategy based on
identifying sentences within a bug report that describe the observable behavior
of a system~\citep{chaparro_observed_2017}, while Misoo et al. explore bug report attachments~\citep{misoo2019}. Rahman et al. observed that  excessive program entities mentioned in the bug report may deteriorate the quality of IR-based bug localization and proposed a query reformulation technique, BLIZZARD~\citep{rahman2018improving}. Le et al. suggests that bug localization
tools can be ineffective for some bug reports and builds a model that can
automatically predict the effectiveness of a IR-based bug localization tool~\citep{le_localization_2017}.
Kim et al. approach the problem similarly, by building a two-phase classifier
that first determines whether the bug report has sufficient information, and, only if it does, recommends a set of code elements~\citep{kim_twophase_2013}.

Changeset-based IRBL has not yet been widely explored. There are two recent publications that
have described such techniques. Locus~\citep{wen2016} is a technique introduced by
Wen et al. that both models and predicts changesets, while Corley et al. explored
training on changesets and predicting code elements using Online LDA~\citep{corley2018}.
Only Corley's technique can be classified as online, in that it does not require
periodic retraining. We improve upon Corley's work in this paper by modeling bug
reports and leveraging the history of fixed bug reports. At the same time,
our approach adapts to the diverse types of content present in bug reports.

While bug localization has generated significant interest among researchers, the lack
of sufficiently large and standardized datasets has hindered clear comparison
among the existing techniques. Recently Bench4BL has made a strong attempt to remedy
this problem by providing a large dataset for bug localization~\citep{lee2018}.
Bench4BL also curates implementations of five existing bug localization techniques
that can be used as baselines. While Bench4BL additionally makes several evaluation
improvements, including finer-grained version matching, it is not completely precise
with respect to time. Such an evaluation would reconstruct the exact snapshot of the
source code when a specific bug report was introduced, instead of
relying on the next release version of the software.

\section{Conclusions and Future Work}\label{sec:conlusion}
In this paper, we take a significant step toward addressing the online bug
localization problem, which is better suited to the rapidity and scale of
modern software development than conventional release-based bug localization. We
motivate the use of online bug localization based on changesets and describe
previously unobserved advantages of this type of model. Our novel bug localization technique, \jitbl, which is based on
changesets and a dual Online LDA model, outperforms a previously reported baseline,
especially when considering different types of bug reports.
Future work includes studying properties of projects that makes them better
suited to bug localization techniques, improving support for projects with short development histories, and leveraging preprocessing techniques that can improve noise filtering in bug reports.

\bibliographystyle{unsrtnat}
\bibliography{paper} 

\end{document}